\documentclass{aa}

\usepackage[flushleft]{threeparttable}
\usepackage{graphicx}
\usepackage{xtab,afterpage}
\usepackage{amsmath}
\usepackage{epstopdf}
\usepackage{url}
\usepackage[flushleft]{threeparttable}
\usepackage{booktabs,fixltx2e}
\usepackage{amssymb}
\usepackage{textcomp}
\usepackage{hyperref} 
\usepackage{multirow} 
\usepackage{times}
\usepackage{txfonts}

\renewcommand{\v}{\ensuremath{\mathbf{v}}}

\def \cesamxx{Cesam2k20\xspace}
\def \eos{EoS\xspace}

\def \K{\mbox{K}}

\makeatletter
\newcommand*{\rom}[1]{\expandafter\@slowromancap\romannumeral #1@}
\makeatother

\title{An impact-free mechanism to deliver water to terrestrial planets and exoplanets}\titlerunning{Delivery of water to Earth from the asteroid belt}
\author{Quentin Kral\inst{1}\thanks{E-mail: quentin.kral@obspm.fr} \and Paul Huet\inst{1} \and Camille Bergez-Casalou\inst{1} \and Philippe Thébault\inst{1} \and Sébastien Charnoz\inst{2} \and Sonia Fornasier\inst{1}}

\institute{LESIA, Observatoire de Paris, Universit{\'e} PSL, CNRS, Universit{\'e} Paris Cit{\'e}, Sorbonne Universit{\'e}, 5 place Jules Janssen, 92195 Meudon, France\\
\and
Université de Paris Cité, Institut de Physique du Globe de Paris, CNRS, 1 rue Jussieu, 75005 Paris, France\\
}

\begin{document}

   \date{Received June 26, 2024; accepted October 3, 2024}



\label{firstpage}


  \abstract
   {The origin of water, particularly on Earth, is still a matter of heated  debate. To date, the most widespread scenario is that the Earth originated without water and that it was brought to the planet mainly as a result of impacts by wet asteroids coming from further out in space. However, many uncertainties remain as to the exact processes that supplied an adequate amount of water to inner terrestrial planets.}
   {In this article, we explore a new mechanism that would allow water to be efficiently transported to planets without impacts. We propose that primordial asteroids were icy and that when the ice sublimated, it formed a gaseous disk that could then reach  planets and deliver water.}
   {We have developed a new model that follows the sublimation of asteroids on gigayear (Gyr)  timescales, taking into account the variable luminosity of the Sun. We then evolved the subsequent gas disk using a viscous diffusion code, which leads to the gas spreading both inwards and outwards in the Solar System. We can then quantify the amount of water that can be accreted onto each planet in a self-consistent manner using our code.}
   {We find that this new disk-delivery mechanism is effective and equipped to explain the water content on  Earth (with the correct D/H ratio) as well as on other planets and the Moon. Our model shows most of the water  being delivered between 20 and 30 Myr after the birth of the Sun, when the Sun's luminosity increased sharply. Our scenario implies the presence of a gaseous water disk with substantial mass for hundreds of millions of years, which could be one of the key tracers of this mechanism. We show that such a watery disk could be detected in young exo-asteroid belts with ALMA.}
   {We propose that viscous water transport is inevitable and more generic than the impact scenario. We also suggest it  is a universal process that may also occur in extrasolar systems. The conditions required for this scenario to unfold are indeed expected to be present in most planetary systems: an opaque proto-planetary disk that is initially cold enough for ice to form in the exo-asteroid belt region,  followed by a natural outward-moving snow line that allows this initial ice to sublimate after the dissipation of the primordial disk,   creating a viscous secondary gas disk and leading to the accretion of water onto the exo-planets.}


   \maketitle
   
\section{Introduction}\label{intro}
 
The origin of water on Earth is still a subject of heated debate that involves many unknowns \citep[e.g.][]{2018SSRv..214...47O}. Water is an essential ingredient for building life \citep[e.g.][]{2014PNAS..11112628M} and understanding its origin would be of paramount relevance. This knowledge could then be extrapolated to  extra-solar systems and help estimate the uniqueness of Earth compared to other exo-planets. In the currently accepted paradigm, it is thought that the inner Solar System (e.g. the Earth's feeding zone) was too warm to hold water ice and that the water had come from the outer parts of the Solar System. The amount of water on terrestrial planets is relatively small; for instance, the Earth containing only 0.023\% water by mass\footnote{taking into account only the hydrosphere, but the Earth may contain between 1 and 10 times this amount, including the water contained in the mantle.}, whereas the giants of the outer Solar System contain up to 40\% water \citep[e.g.][]{2014MNRAS.441.2273H}. 

The most widespread theory to date is that the water may have been brought in by impacts from icy bodies, mainly asteroids and possibly comets in smaller quantities \citep[e.g.][]{2006Icar..183..265R,2019AJ....157...80C}. One key constraint supporting this scenario is that Earth's D/H ratio is very similar to that of carbonaceous chondrites \citep{2012Sci...337..721A}, which are thought to have originated from C-type asteroids, suggesting that they were Earth's main source of water. There are several different scenarios  to explain how outer icy bodies could have been brought to the inner Solar System, but they all share some common features, which can be summarised as follows. Some icy asteroid-like bodies from the outer regions of the Solar System (4-10 au for the most part) were perturbed (either by scattering induced by long term resonances or by more violent instability), allowing them to populate the asteroid belt and, at the same time, cause impacts on  terrestrial planets. As these bodies are icy, they can then supply water to the planets via these impacts \citep[e.g.][]{2006Icar..183..265R,2011Natur.475..206W,2017Icar..297..134R}. 

However, some alternative scenarios suggest that the Earth's water could be local in origin and may have been inherited very early in the Earth building blocks from a material similar to enstatite chondrites, which are very dry (but can contain enough hydrogen to provide sufficient water) and very similar to the material that formed the Earth \citep{2020Sci...369.1110P}. This may, however, be in disagreement with the composition of the Earth's mantle and core, where models suggest that water is added to the Earth mainly at the end of its formation \citep{2008GeCoA..72.1415W,2015Icar..248...89R}. It should also be noted that there is evidence that water could have been present on Earth rather early, probably $<150$ Myr after the Sun's birth, based on observations of very old zircons \citep{2001Natur.409..178M,2001Natur.409..175W}, which leads to some wide interval for the water arrival.

However, most of the data suggest that the water was delivered before the late veneer, which corresponds to the latest stages of planetesimals impacts, occuring after the Moon-forming impact that is dated between 30-200 Myr \citep{2009GeCoA..73.5150K}, but probably closer to 50-60 Myr \citep{2017SciA....3E2365B} after the birth of the Sun. The late veneer is expected to have contributed an additional $5 \times 10^{-3}$ M$_\oplus$ of chondritic material based on measurements of highly siderophile (iron-loving) elements in the Earth's mantle \citep[e.g.][]{2015GMS...212...71M}. However, at most 10\% of the water may have been delivered to the Earth by impacts during late veneer according to geochemical and isotopic arguments \citep[e.g. noble gas budget,][]{2013GeCoA.105..146H,2015GMS...212...71M} and impact models showing that the number of impacts needed to explain the mass of terrestrial water would lead to an atmosphere that is too massive compared with the current atmospheric mass \citep{2020MNRAS.499.5334S}. It is therefore likely that most of the water had already been delivered before $\sim$60 Myr.

Although impact scenarios are currently favoured to explain the delivery of water to planets, they are in fact highly contingent. Indeed, in most of these impact scenarios, it takes a complex dynamic history (e.g. resonances between planets) to move icy bodies from the outer regions of the Solar System towards the planets at the right time. Most of these scenarios require fine-tuning in one way or another. It would be interesting to find a more universal mechanism and this is the motivation for this work.

In this paper, we propose studying a new mechanism that could provide water to planets without impacts. The main idea is that the primordial asteroid belt must have contained icy asteroids, given that most C-type asteroids (which represent the largest fraction of asteroids in the main belt) are hydrated. This water ice would have sublimated in a few tens of Myr, creating a disk of gas that could then spread towards the inner planets and bring them water.

Therefore, our mechanism relies on the asteroid belt composition and mass and there are several possibilities. The young asteroid belt may have been born massive, with a mass up to a thousand times greater \citep[e.g.][]{1980ARA&A..18...77W} than its current mass of $\sim 4 \times 10^{-4}$ M$_\oplus$ \citep[e.g.][]{2018AstL...44..554P}, according to the MMSN model \citep{1981PThPS..70...35H}. It could also have been born with a low mass and subsequently became populated by bodies from the outer region of the Solar System \citep[e.g.][]{2017Icar..297..134R}. The current asteroid belt is composed of two main classes of asteroids, with S-class objects being most common in the inner main belt and C-class objects dominating the outer main belt, with substantial mixing between the two populations \citep[e.g.][]{2013Icar..226..723D}. C-types are spectroscopically related to carbonaceous chondrites, which generally contain $\sim$10\% water by mass \citep{2018SSRv..214...36A}. As we show in more detail in the next section (Sect. \ref{whywaterice}), young C-types may have been rich in water ice and the young asteroid belt would have had a fairly high sublimation rate, building up a disk of water gas.

It is even possible that disks of gaseous water exist outside of our Solar System, in exo-planetary systems. It would then build up from gas released in exo-asteroid belts. However, those water disks have not yet been detected but it may well be in the close future. Indeed, it would then become similar to exo-Kuiper belts, which are the equivalent of the Solar System's Kuiper belt around other stars, that are now observed to be releasing significant quantities of gas, in particular CO and carbon \citep[e.g.][]{2017ApJ...849..123M,2023ApJ...951..111C}. This phenomenon was not predicted by previous models of planetary formation and has only recently been discovered \citep[e.g.][]{2013ApJ...776...77K}, but it shows that CO ices are present in the outer regions of extra-solar systems and that they can sublimate and produce disks of gas. These disks can propagate towards exo-planets and deliver them gaseous CO, carbon, or oxygen in a relatively efficient late accretion process \citep{2020NatAs...4..769K}. 

The closest observation to a disk of water in an exo-planetary system comes from {\it Spitzer} observations of HD 69830 (a star not very different from our Sun, with an asteroid belt at $\sim$ 1 au), showing that water ice may be present in its exo-asteroid belt and that it should be sublimating, building up a disk of gas \citep{2007ApJ...658..584L}. Gaseous water could also spread towards the inner regions of exo-planetary systems and provide water for exo-planets in the system's habitable zone. This is a new mechanism that could be universal across all planetary systems and  we explore it in this paper from a Solar System perspective. We  show that current facilities, such as ALMA, may be able to detect these disks of gas and thus test this new theory using extra-solar system data.

\section{Water ice in the young asteroid belt}\label{whywaterice}

The foundation of our model is that young asteroids were icy. This   hypothesis, which is the basis of this paper may not have been examined in sufficient detail, even in the case of previous (impact) models working on a similar hypothesis.

When we look at asteroids today, they appear to be mostly free of water ice, with the exception of Ceres, the largest asteroid, where exposed water ice was detected locally \citep{2016Sci...353.3010C,2017LPI....48.2447P,2018SciA....4.3757R}, and potentially Themis \citep{2010Natur.464.1322R,2010Natur.464.1320C} as well as (65) Cybele \citep{2011A&A...525A..34L}, large $\sim$ 200 km bodies in the outer asteroid belt. The dwarf planet Ceres contains $\sim 20\%$  water ice in the near-surface \citep{2017Sci...355...55P}, and the bulk crustal average water content is estimated to be greater than 60 vol.\% \citep{2017E&PSL.476..153F,2020NatAs...4..748P}. Moreover localised sources of water vapour have been detected \citep{2014Natur.505..525K}. In addition, Ceres appears to have an icy mantle and may even have an ocean of water beneath its surface \citep{2020AsBio..20..269C}. Therefore, from current observations it appears that water ice is only detected on a few of the larger asteroids in the outer parts of the asteroid belt. However, we note that ice on the surface is short-lived because the totality of the asteroid belt is located within the snow line (the distance at which water transitions from ice to gas; see Sect.~\ref{universal} for a detailed discussion) so that the ice reservoir should come from further down. Moreover, it is now known that water ice is not easy to detect on surfaces of dark objects. Indeed, in situ missions such as Rosetta clearly show that even on comets, it is difficult to be observe ice on the surface. For 67P/Churyumov-Gerasimenko, \citet{2023A&A...672A.136F} showed that 67P/CG surface is dominated by refractory material and that the detected exposed water ice during the Rosetta mission represents only 0.1\% of the nucleus surface. Additionally, the characteristic size of exposed chunks of water ice on 67P/CG is of the order of 1 m$^2$ or less, on average, implying that high spatial resolution is required to detect it. However, there are several compelling arguments that indicate that even if it is difficult to observe ice on the surface, water ice is abundant just below the top refractory-dominated layer,  at a depth of 10 cm to 1 m, as can be seen after cliff collapses and scarps formation \citep[e.g.][]{2017NatAs...1E..92P,2017MNRAS.469S..93F,2019A&A...630A...7F} or by the imprint of Philae on a boulder \citep{2020Natur.586..697O}. 

Even if  ice is rare on the surface of asteroids, the observations of meteorites show that some of them are full of hydrated minerals. Given that meteorites observed today do not have water ice, their water content is defined as the amount of water that could be reformed after heating the solids; thus, it goes on to become a gas mixture, where H and O are assumed to be recombined to potentially form water. There is a clear segregation between the inner belt, dominated by water-poor S-type asteroids and the outer belt dominated by water-rich C-type asteroids \citep[e.g.][]{1985GeCoA..49.1707K,2018SSRv..214...36A}. The water fraction in S-type asteroids is as low as 0.1\%, whereas that of C-type asteroids is around 10\%. It is therefore generally thought that most of the water delivered to Earth must come from C-type asteroids. We note that water in asteroids  currently observed  today is in the form of hydrated minerals and not in the form of ice. However, it is important to stress that the composition of today's asteroids could be very different from their make-up in the early Solar System. It is indeed possible that most asteroids (at least the C-types rich in hydrated minerals) were initially composed of water ice that sublimated over the 4.6 Gyr of evolution of the Solar System. Several pieces of evidence point in this direction:

First, the study of meteorites show that carbonaceous chondrites (which are very likely to  have C-type asteroids as parent bodies) present evidence of fluid-rock interactions because they host numerous aqueously formed minerals such as phyllosilicates, carbonates, and magnetite \citep{2006LPI....37.2074B}. Clays produced by the effect of water \citep{1989E&PSL..95..187A} are also found on chondrites. It is thought that aqueous alteration may have been driven primarily by impact on water ice, rather than by radiogenic heating given the correlation of aqueous alteration with petrofabric strength \citep[most likely arising due to shock deformation; e.g.][]{2021GeCoA.299..219S}.

The second argument comes from the study of asteroids using telescopes or in situ missions. From observations of a great number of C-type asteroids, \citet{2014Icar..233..163F} show that aqueous alteration becomes important beyond 2.3 au in the asteroid belt. Moreover, \citet{2015AJ....150..198R} show that 70\% of C-type asteroids are hydrated. Recent in situ missions targeting two carbonaceous asteroids Bennu (OSIRIS-REx mission) and Ryugu (Hayabusa 2 mission) confirmed the presence of hydrated minerals on both \citep{2019NatAs...3..332H,2019Sci...364..272K} and of organic and carbonate material on Bennu \citep{2020Sci...370.3522S}. From the samples returned to Earth, measurements in the lab found even deeper lines, namely, even more hydrated minerals, than what could be observed from the spacecrafts \citep{2019NatAs...3..332H, 2021Icar..36314427P, 2022LPICo2695.6226P, 2024LPICo3040.1366H}. Bennu and Ryugu are near-Earth asteroids with semi-major axes close to 1 au, but originating from the inner main belt, around $2.1-2.5$ au \citep{2015Icar..247..191B,2018PEPS....5...82W}. Ryugu's isotopic analysis and paleomagnetic studies date the onset of fluid activity and hence aqueous alteration between $<$1.8 and 6.8 Myr after the formation of the CAI \citep{2023NatAs...7..309M,2024E&PSL.62718559M}.

Thirdly, there are numerous observations of so-called main belt comets or activated asteroids, which are asteroids showing signs of activity which may be comet-like \citep{2006Sci...312..561H, 2022arXiv220301397J}. Several of these main belt comets come from the Themis family \citep{2008Icar..193...85N, 2012ApJ...748L..15H} in the outer main belt. \citet{2016Icar..269....1F} studied some members of the Themis family and found spectral features near 3 $\mu$m that could be associated with hydrated minerals or water ice. These features could be confirmed in the future using JWST.

Fourth, and perhaps more crucially, models show that the water snowline that currently lies beyond the asteroid belt was much closer in the past, allowing water ice to form throughout the asteroid belt (see Sec.~\ref{universal} for a simplified model of the snowline). The snowline occurs at temperatures of around 145-170 K \citep[][]{1981PThPS..70...35H,2003ApJ...591.1220L} and recent simulations of proto-planetary disks show that the snowline moves rapidly inwards to stabilise around 2 au after a few million years \citep{2015A&A...577A..65B} and only changes when the disk dissipates to a new position beyond 3 au. It is even possible that the snowline moved inwards to less than 2 au as accretion slowed and the disk cooled and then got stuck (or fossilised) around 2.3-2.7 au \citep{2016Icar..267..368M}, which corresponds to where the separation between water-rich and water-poor asteroids occurs in the current belt \citep[e.g.][]{2000orem.book..413A,2000M&PS...35.1309M}. Fossilisation would occur in the proto-planetary disk phase because of Jupiter blocking the flow of pebbles as it reaches the isolation mass, so that even if the inner main belt becomes cooler, it cannot be filled with icy material anymore and water ice never comes back on those asteroids \citep{2016Icar..267..368M}. Interestingly, this scenario predicts that extra-solar systems without giant planets should be much richer in water in their inner regions. Another possibility is that the  magneto-rotational instability (MRI) was not active in the inner regions of the belt; furthermore, due to the resulting dead zone, matter can accumulate beyond the dead zone, becoming more massive and reaching higher temperatures (because of viscous heating), thereby pushing the snowline further out than in a fully MRI turbulent model \citep{2012MNRAS.425L...6M}.

Fifth, most current scenarios for the transport of water to the planets are based on the injection into the asteroid belt of planetesimals from the outer regions (4-10 au) and their accretion onto the planets at the same time due to intense scattering \citep[e.g.][]{2006Icar..183..265R,2011Natur.475..206W,2017Icar..297..134R}. These wet asteroids are considered to quantify the amount of water that can be delivered to Earth, but studies have never consider that wet asteroids implanted in the main belt will be icy and their subsequent evolution has been omitted. The aim of this paper is to rectify this gap and explore whether sublimated ice in the main belt is an effective means of transporting water to planets that could dominate over impacts.

Sixth, another way to learn more about ice in the asteroid belt is to study extra-solar systems with warm exo-asteroid belts. The water content can be studied at the beginning of the proto-planetary disk phase in the inner few au and the extent of the water snowline can be explored at high resolution \citep{2024NatAs.tmp...49F}. Indeed, various observations of water vapour have found cold and warm water lines in some young disks by {\it Herschel} \citep{2021A&A...648A..24V}, {\it Spitzer} \citep{2010ApJ...722L.173P}, JWST \citep{2023ApJ...957L..22B}, and also from the ground \citep{2015ApJ...810L..24S}. This is not surprising as water vapour is expected to be released near the snowline as dust or pebbles migrate into the disk interior, which is an important prediction of disk models during planet formation. What is even more interesting is that water ice is detected at a later stage, at the debris disk stage, when the star is on the main sequence and the disk resembles our asteroid belt. Indeed, in the system around the K0V-type star HD 69830, aged $\sim$ 3-10 Gyr, there is an exo-asteroid belt close to 1 au (making it a close analogue of our Solar System), where water ice is detected on dust grains produced by collisions in the warm belt \citep{2007ApJ...658..584L}. This is somewhat surprising because the system is old and the temperature at 1 au is higher than the sublimation temperature of water. However, it could be that water is sublimated in this belt, which creates a certain opacity to incident light and does not allow for much heat to reach the exo-asteroids, making them cold enough to keep their water ice. We note that it is likely that many systems with exo-asteroid belts should show the presence of water ice, particularly in the youngest systems.

\section{The model}\label{theeq}

We developed a new model  that is able to follow the sublimation of ice in the asteroid belt from its youth to Gyr timescales accounting for the varying luminosity of the Sun. The gas released from sublimation then makes up a gas disk composed of water, which evolves viscously within the frame of our model and spreads radially inwards and outwards towards the planets. We then follow the water gas mass that gets accreted onto the terrestrial planets for two different scenarios: 1) the asteroid belt was initially massive, as indicated by the MMSN model, and depletes early at $\sim 50$ Myr, 2) the asteroid belt started with its current mass. We describe those two scenarios in detail and what effects they correspond to  in the next section.

\subsection{Mass, spatial distribution and size distribution of the asteroid belt}\label{sizedis}

In this paper, we explore two scenarios that lead to different initial belt masses. First, we modelled a belt of similar mass to today's, which has only undergone very minor collisional evolution, which would be the case if it had started with the current mass of $\sim 4 \times 10^{-4}$ M$_\oplus$ \citep[e.g.][]{2018AstL...44..554P}. This scenario is consistent with the likely end result of the ``Grand Tack'' \citep{2011Natur.475..206W} or the ``empty belt model'' \citep[which gets refilled by planet interactions,][]{2017SciA....3E1138R}. 

The second scenario is based on a more massive belt that gets depleted to current asteroid belt mass level after a time, $t_{\rm dep}$. The depletion is implemented as a step function reduction in the belt mass after $t_{\rm dep}$. In our fiducial model, we used an initial rocky mass of 0.1 M$_\oplus$ and $t_{\rm dep}=50$ Myr. This is a generic scenario assuming that the belt started with a mass similar to the MMSN model and it got emptied over time, either via an instability or dynamical depletion due to resonances with a planet, such as Jupiter; however, we note that in this latter case, the depletion would be smoother than assumed here, but our results would not change much (as discussed later in this paper).

The massive case scenario is consistent with the early instability model \citep[that can deplete the asteroid belt at the 99-99.9\% level,][]{2018Icar..311..340C,2019AJ....157...38C}, which is an updated and amended version of the ``Nice model'' \citep{2005Natur.435..466G} for which the instability occurs much earlier between 10-100 Myr. Some newer studies have even asserted that the instability may have occurred between 60 and 100 Myr \citep{2024Sci...384..348A}, but this is still strongly debated, as explained in \citealt{2024arXiv240410828I}. We note that our model is generic and not limited to an early instability as an instability may not be necessary to deplete the initially massive asteroid belt. Indeed, natural processes such as mean motion resonances with Jupiter \citep{2006Icar..183..265R} or the $\nu_6$ secular resonance with Saturn at 2.1 au \citep{2000M&PS...35.1309M} would be able to clear the belt over long timescales. This clearing can operate over less than one Gyr and deplete more than 99\% of the belt \citep{2001Icar..153..338P}.

We emphasise that the primordial asteroid belt may have been even more massive; according to previous works, it could have reached $\sim 0.5$ M$_\oplus$ \citep[e.g.][]{2000M&PS...35.1309M,2001Icar..153..338P}. Furthermore, simulations accounting for self-gravity or planet stirring find an upper limit of 2 M$_\oplus$ \citep[e.g.][]{2007Icar..191..434O,2018ApJ...864...50D,2019AJ....157...38C}.

\begin{figure}
   \centering
   \includegraphics[width=9.cm]{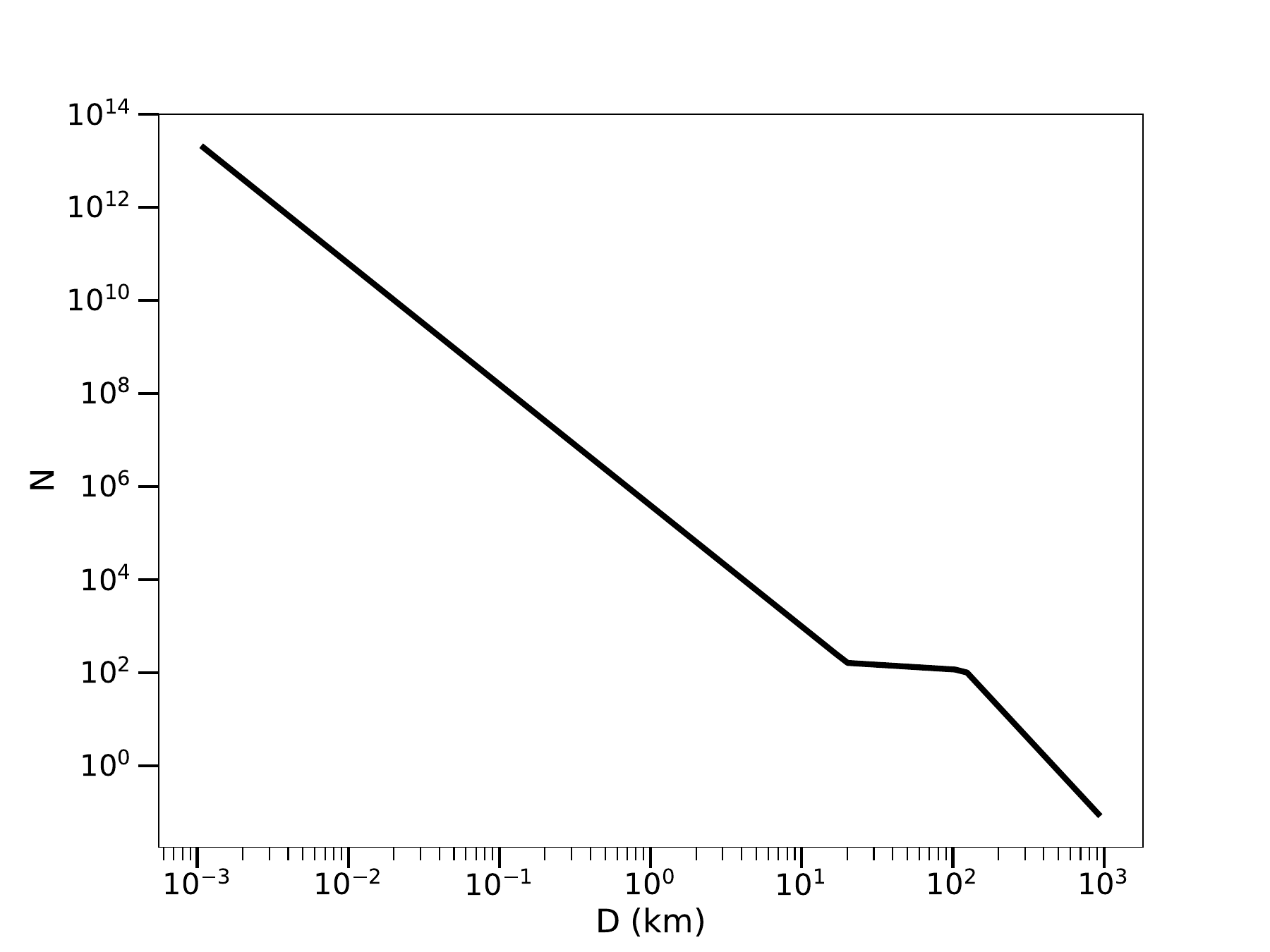}
   \caption{\label{figpsd} Particle size distribution of bodies in the asteroid belt used in our model represented by the number of bodies in each diameter bin. There are three regimes with different slopes as described in the main text.}
\end{figure}

We base our fiducial model on the characteristics of the spatial and size distributions of the current asteroid belt, which we  describe here. The main belt extends from $\sim$2 to $\sim$3.3 au (between the strong 4:1 and 2:1 resonances creating Kirkwood gaps). These are the values we use in our model to define the radial extent of the asteroid belt. The radial distribution of the early asteroid belt is defined using the MMSN model such that the surface density varies as $r^{-3/2}$. The size distribution we use is similar to that of the current asteroid belt with three power laws to define three regimes between different solid body diameters: $D_{\rm min}=1$ m, $D_{\rm med}=$20 km, $D_{\rm big}=$120 km, and $D_{\rm max}=1000$ km, with slopes $q_{\rm low}$ for $D_{\rm min}<D<D_{\rm med}$, $q_{\rm med}$ for $D_{\rm med}<D<D_{\rm big}$, and $q_{\rm high}$ for $D_{\rm big}<D<D_{\rm max}$. In our fiducial simulation, we use $q_{\rm low}=3.6$, $q_{\rm med}=1.2$, and $q_{\rm high}=4.5$ (Bottke et al. 2005). This is represented in Fig.~\ref{figpsd}, where we note $N$ the number of solid bodies in each size bin. Later in this paper, we discuss the fact that given the very long collisional timescales of the largest bodies in the belt, the size distribution is not evolving much during the lifetime of the Solar System.

Based on observations, the separation between wet and dry asteroids is around 2.3 au (\citealt{2014Icar..233..163F}, or up to 2.7 au, see the fossilised snow line in \citealt{2016Icar..267..368M}) and we therefore initially only place ice between 2.3 and 3.3 au in our model. Because there are no signs of hydration below 2.3 au, we consider that asteroids there may have never hosted water ice.

\subsection{Composition of primordial asteroids}

Many arguments for the presence of water ice in the young asteroid belt were already provided in Section \ref{whywaterice}. According to models, the snow line in the Solar System has fallen below 2 au \citep[e.g.][]{2015A&A...577A..65B} in the proto-planetary disk phase. Thus,  asteroids would have been cold enough in this young disk phase ($<5$ Myr) to contain water ice, which would be released as gas later when the young disk dissipated and the snowline moved outwards when entering the debris disk phase. Indeed, as the snow line is now beyond the asteroid belt \citep[e.g.][]{2016Icar..267..368M}, this ice should have sublimated and should no longer be found there, if it has been sublimated within the age of the Solar System. In Section \ref{universal}, we show a simple model of the temperature evolution from the proto-planetary disk phase to the debris disk phase, which explains rather straightforwardly how the snow line moves between the different phases.

\citet{2003ApJ...591.1220L} gives a theoretical estimation of the fraction of water that may condense beyond the snowline in an MMSN model and find that we should expect $\sim$ 50\% water by mass, which is indeed similar to comets \citep{2015SSRv..197...47B}. As for asteroids, it is clear that liquid water was present initially given the presence of hydrated asteroids as liquid water needs to interact with rock to create the hydrated minerals \citep{2015NatCo...6.7444D, 2017GeCoA.213..271V}. Moreover, liquid water is not stable on the surface of asteroids due to the low gas pressures there and the original form of water would have been ice. Indeed, the most likely cause of the presence of liquid water is the melting of water ice following impacts \citep[e.g.][]{2021GeCoA.299..219S}. 

To investigate how much water might have interacted with minerals, oxygen isotopes can be used to obtain initial water/rock ratios. Using this technique, \citet{2018E&PSL.482...23M} find that the water/rock ratio can vary between 1\% and 40\% depending on the type of carbonaceous chondrites (increasing in that order CO, CV, CR, and CM), showing some heterogeneity in the liquid water content. This gives an idea of the potential initial ice content that must have been greater than the liquid content. 

However, there are no direct connection between the liquid and icy content and it is very difficult to determine the initial amount of water ice (that has since sublimated), as it leaves no trace on the surface of asteroids or meteorites. As explained below, one way of observing the early sublimation of water ice would be to study extra-solar systems and check whether this phase of water gas release is common. Another way would be to study the effect of the released gas on planet formation, as we have done in this article. For our study, given the uncertainties, we leave the amount of water ice that can sublimate as a free parameter lower than 50\%, but for our fiducial model, we assume that the asteroids initially contained $f_{\rm ice}=20$\% water by mass.

This leads to an ice mass between 2.3 and 3.3 au of $\sim 6.4 \times10^{-5}$ M$_\oplus$ for scenario 1 (low mass) and $\sim 1.6 \times 10^{-2}$ M$_\oplus$ for scenario 2 (high mass). The ice mass is added in addition to the rocky mass in our model so that we get a total solid mass of $\sim 4.6 \times10^{-4}$ M$_\oplus$ for the first scenario and $\sim 0.12$ M$_\oplus$ for the second.

\subsection{Sublimation of asteroid ices}

The temperature of solid bodies after the dissipation of the proto-planetary disk is assumed to be a black body temperature so that $T^4_{\rm bb}=L_\star (1-A)/(16 \pi \sigma r^2)$ where $L_\star$ is the solar luminosity, $\sigma$ is the Stefan-Boltzmann constant, $A$ is the albedo, which we take to be equal to 0.06 \citep[because C-type asteroids are very dark\footnote{Note: the current Bond albedo of C-type asteroids may even be lower by a factor 3 \citep{1989aste.conf..557H,2022PSJ.....3...95V} but it was probably larger in the past.},][]{2023A&A...680A..10D}, and $r$ is the distance of the body from the Sun. We note that the young Sun after $\sim 3$ Myr was much less luminous than it is today. We  follow the evolution of luminosity as a function of time $L_\star(t)$ with a state-of-the-art model in Section \ref{sun}. Finally, we obtain: 
\begin{equation}\label{Tbb}
T_{\rm bb} \sim 278 \, {\rm K} \left( \frac{L_\star(t)}{L_\odot} \right)^{1/4} \left( \frac{r}{1 {\rm au}} \right)^{-1/2} \left(1-A \right)^{1/4}.
\end{equation}

\noindent This leads to a temperature range of 130-170 K for the main belt at the beginning of the simulation when the solar luminosity is $\sim 0.6$ L$_\odot$ increasing to 150-190 K for 1 L$_\odot$. This means that sublimation can be efficient from the very beginning given that the sublimation temperature of water ice is around 145-170 K. For simplicity, we use Eq.~\ref{Tbb} for the temperature $T$ of both solids and gas. We will discuss later how a different gas temperature would change our results.

We assume that the sublimation rate per unit surface area of water ice $Z$ for an icy body in vacuum is given by 
\begin{equation}\label{Zeq}
Z(T)=\frac{a_1}{\sqrt{T}}\exp(-\frac{a_2}{T}) \, {\rm [in \, m}^{-2}{\rm s}^{-1}], 
\end{equation}

\noindent where $a_1=7. 08 \times 10^{35}$ m$^{-2}$ s$^{-1}$ K$^{1/2}$ and $a_2=6062$ K \citep{1991Icar...90..319L}. Therefore, the solid mass loss rate for a solid body of radius $a$ is given by 

\begin{equation}\label{mdoteq}
\dot{m}=-4\pi a^2 \mu_w m_p Z(T),
\end{equation}

\noindent where $\mu_w=18$ is the mean molecular weight of the water molecule and $m_p$ is the mass of a proton. We note that we keep $a$ constant over the whole simulation. We remain agnostic concerning the exact distribution of ice with respect to rock (e.g. mixed or in an upper layer). In fact, Eq.~\ref{Zeq} works either for the case where the entire surface of the asteroid is covered in ice, or for a mixture of rock and ice, on the assumption that the ice is sublimated at depth and the vapor produced rises rapidly to the surface. As will be clear later on, we keep track of how much ice mass has been lost over time for each asteroid, and we stop producing water once the ice reservoir has been exhausted.

We can then estimate the survival time of a certain mass of ice $M_{\rm ice}$ on an icy body of size $a$ and density $\rho$ such that 
\begin{equation}
t_{\rm ice}=\frac{M_{\rm ice}}{\dot{m}}=\frac{a \rho}{3Z(T) \mu_w m_p},
\end{equation}

\noindent which is close to a Gyr for the largest bodies ($\sim 500$ km) in the asteroid belt and, for example, $10^5$ times smaller for a 5 m size body, that is 10,000 yr. This means that the lower end of the size distribution will sublimate its water content quite quickly (and even more so near the Sun) but that there will still be a long-lived water vapour input from the sublimation of the larger bodies that will diminish with time but which can last for hundreds of Myr, as we observe in our simulations.

In the formalism of \citet{1991Icar...90..319L}, the equilibrium vapour pressure is given by 
\begin{equation}
P_{\rm eq}=Z(T) \sqrt{2 \pi \mu_w m_p k_b T},
\end{equation}

\noindent where $k_b$ is Boltzmann's constant and the gas vapour pressure can be calculated from the surface density of water gas $\Sigma_w$ as 
\begin{equation}
P_{\rm vap}=\frac{\Sigma_w k_b T}{2H \mu_w m_p}.
\end{equation}

When $P_{\rm vap}<P_{\rm eq}$, using Eq.~\ref{mdoteq}, we convert the water ice to steam until we reach $P_{\rm vap}=P_{\rm eq}$. In all subsequent simulations, the ice mass is not sufficient to reach $P_{\rm eq}$ and we convert the lost ice mass into vapour every time step and for each size bin, thus increasing $\Sigma_w$. When $P_{\rm vap}>P_{\rm eq}$, the water vapour condenses to form solid ice. In this case, excess water vapour is removed from the gas phase to maintain equilibrium vapour pressure and ice is returned to the smaller dust by adding the contribution $\Delta \Sigma_{\rm cond}=(P_{\rm vap}-P_{\rm eq}) 2H \frac{\mu_w m_p}{kT}$ and removing an equal amount from the vapour phase. However, the case where $P_{\rm vap}>P_{\rm eq}$ does not occur in the simulations we have carried out for the Solar System.

\subsection{Viscous evolution of gas released from asteroids}

At each time step, we make the gas evolve viscously using the following equation \citep{1974MNRAS.168..603L}:

\begin{equation}\label{eqdiff}
 \frac{\partial \Sigma}{\partial t}=\frac{3}{r} \frac{\partial}{\partial r} \left[ \sqrt{r} \frac{\partial}{\partial r}(\nu \Sigma \sqrt{r})  \right] + \dot{\Sigma}_{\rm io} 
 ,\end{equation}

\noindent where $\nu$ is the kinematic viscosity, $\Sigma (r,t)$ is the surface density of the gas, $r$ is the radial variable, and $\dot{\Sigma}_{\rm io} (r,t)$ is the rate of input/output to the surface density at radius $r$ and time $t$ which corresponds to the gas sublimated by asteroids or recondensed on dust. 


In practice, we calculate how much mass $m^+(a)=N(a)\dot{m}(a) \Delta t$ is lost to vapour by sublimation for the different size bins of radius $a$ containing $N(a)$ solid bodies, over one time step and at a specific radius $r$ in the belt, and how much is recondensed $m^-$, to finally sum over all size bins to obtain:

\begin{equation}
\dot{\Sigma}_{\rm io}(r)= \frac{\sum_{a_{\rm min}}^{a_{\rm max}} m^+(a)-m^-}{\pi [(r+\delta r)^2-r^2] \Delta t} \sim \sum_{a_{\rm min}}^{a_{\rm max}} \frac{N(a)\dot{m}(a)}{2 \pi r \delta r},
\end{equation}

\noindent where $\Delta t$ is the time step of the simulation, $\delta r$ is the width of the radial bin and $a_{\rm min}$ and $a_{\rm max}$ are the mininum and maximum bin sizes considered (see Sect. \ref{sizedis}). 

For the viscosity $\nu$, we use the standard $\alpha$-parameterisation of \citet{1973A&A....24..337S}, such that 
\begin{equation}
\nu = \alpha c_s H,
\end{equation}

\noindent where $c_s=\sqrt{k_b T_g/(\mu m_p)}$ is the speed of sound, $T_g$ is the gas temperature, $\mu$ is the molecular weight of the gas mixture (which may be different from $\mu_w$), and $H=c_s/\Omega$ is the local scale height of the disk, with $\Omega$ the orbital frequency. We use a fiducial value of $\alpha=10^{-2}$ in the rest of the paper and discuss other values in Sect.~\ref{alphavar}.

To evolve Eq.~\ref{eqdiff} forward, we use a finite difference scheme with an explicit spatial integration and implicit temporal evolution. In practice, we numerically integrate the differential equation using the python function {\it solve\_ivp} from {\it scipy.integrate} with the integrator LSODA and an adaptive timestep.

\subsection{Shielding, ionisation, and dissociation of water}\label{secshiel}

Initially, when water ice sublimates into the gaseous phase, it is released as a water molecule, H$_2$O. However, under the effect of UV photons (from the star and the interstellar medium), water can dissociate or ionise over fairly short periods, of the order of 1 to 10 days, at the location of the asteroid belt. However, if enough water molecules are present, \citet{2009Sci...326.1675B} have demonstrated that water can protect itself against photons. This water shielding mechanism not only protects water molecules, but also shields a wide range of wavelengths, as ozone does in the Earth's atmosphere, which can allow other molecules (if present) to survive. This is a process known as water UV shielding. 

The absorption cross-section of a water molecule is $\sigma_w=5 \times 10^{-18}$ cm$^2$ between 91.2 and 200 nm, which is similar to that of OH \citep{2009Sci...326.1675B}. To reach a water column density $N_w$ such that the optical depth $\tau_w$ is 1, we need $N_w=\tau_w/ \sigma_w=1/\sigma_w$ and then the water starts to protect itself from dissociation. This corresponds to $N_w=2\times 10^{16}$ cm$^{-2}$ or a critical surface density $\Sigma_{\rm crit\_water}=N_w \mu_w m_p=6\times 10^{-6}$ kg/m$^2$ (given that the shielding operates vertically but we note that some radial shielding may also be at work). This is somewhat similar to carbon shielding in exo-Kuiper belts \citep[except that C shields CO in this case,][]{2019MNRAS.489.3670K} for which the cross-section is a factor of 3 larger with $\sigma_c=1.6 \times 10^{-17}$ cm$^2$. The survival time scale of water can then be approximated as $t_{\rm shield} = t_w \exp(\sigma_w N_w)$, where $t_w$ is the standard survival time scale of unshielded water (which would be fixed by the stellar flux when radial shielding is not strong and by the interstellar medium flux otherwise). This is well known in the proto-planetary disk community where gaseous water is observed in such quantities that shielding is necessary \citep{2007prpl.conf..507N}.

For the case where there is no shielding, we use the photo-dissociation and photo-ionisation time scales given by \citet{2015P&SS..106...11H} at 1 au where we give the values as a range to take into account periods where the Sun is faint or active\footnote{One can convert to timescales at x au by multiplying the values by x$^2$.}:

\begin{itemize}
\item For H$_2$O - the dominant photo-dissociation channel is H+OH (0.66-1.13 days): photo-ionisation leads to H$_2$O$^+$ (14.0-35.0 days);
\item For OH - the dominant photo-dissociation channel is O(3P)+H (1.62-1.77 days according to theory, or 0.85-0.93 days from experiment);
\item For O(3P) - photo-ionisation leads to O$^+$ (17.6-47.5 days);
\item For H -  photo-ionisation leads to H$^+$ (67.4-160 days).
\end{itemize}

We note that in simulations starting with a high mass of solids (scenario 2), we have a case where water is self-protecting and can accumulate. This aspect becomes clear later on when we present our results. Otherwise, water will photo-dissociate to O and H atoms and some of it could photo-ionise and produce electrons. Therefore, here we consider a mean molecular weight $\mu=18$ for the high mass cases (scenario 2), and $\mu=6$ otherwise (because of the mixture of atomic O and H, scenario 1). We note that we fixed $\mu$ to its initial value for the whole simulation.

\subsection{Gas accretion onto planets}\label{modpla}

As the gas flow radially spreads inwards to the Sun and outwards to the outer planets of the Solar System, some of the gaseous material interacts with the planets by entering their Hill spheres. To model this process, we add a sink cell for each planet to calculate the mass accreted $M_{\rm acc}$ onto them. We calculated the radial mass flux through each cell by computing\footnote{This is because $\dot{M}_r=-2 \pi r v_r \Sigma$ and the radial velocity $v_r=-3/(\Sigma \sqrt{r}) \partial/\partial r (\nu \Sigma \sqrt{r})$ \citep{1974MNRAS.168..603L}.} $\dot{M}_r=6 \pi \sqrt{r} \frac{\partial}{\partial r}(\sqrt{r}\Sigma \nu)$ at the locations $r_p$ (0.38, 0.72, 1, and 1.52 au) of the four terrestrial planets (Mercury, Venus, Earth, Mars) and multiplying by the time step $\Delta t$. We then calculated the Hill radii $R_h=r_p (M_p/(3M_\odot))^{1/3}$ of the planets (using masses $M_p$ of 0.0553, 0.815, 1, and 0.107 M$_\oplus$) and compared the scale height of the gas disk $H$ to $R_h$. For the cases where $H>R_h$ (so part of the gas mass cannot be accreted by the planet\footnote{We note that one could think that in addition to this vertical correction there could be an azimuthal correction. However, we assume that the planets have time to orbit several times before the accreted gas can cross radially, so that no azimuthal correction factor is needed (because the viscous time scale is very large compared to the orbital time scale).}), we multiplied $M_{\rm acc}$ by a factor $R_h/H$ \citep[see details in][]{2020NatAs...4..769K}. For instance, at the beginning of the simulations, $R_h/H$ equals 0.24, 0.49, 0.49, and 0.20 for Mercury, Venus, Earth, and Mars, respectively. We note that $H$ scales as $T_g^{1/2}$, which increases in our model with $L_\star$, so that $H \propto L_\star^{1/8}$ and $R_h/H$ can at most be divided by a factor 1.05 for a solar luminosity.

We also take into account the fact that 3D hydro simulations show that some gas can pass through or circulate back to where it comes from without being able to be accreted onto the planets by multiplying $M_{\rm acc}$ by a factor $f_{\rm hydro}=1/2$ \citep[e.g.][]{1999MNRAS.303..696K,2018A&A...617A..98R,2020A&A...643A.133B}. Finally, we use the work of \citet{2020NatAs...4..769K} to calculate the efficiency of planets to accrete gas once this gas has reached the Hill sphere. We find that the planets can cool fast enough relative to $\dot{M}_r$ for all of the incoming gas. As a consequence, we assume that all incoming gas that can be accreted is accreted and fix the cooling efficiency parameter $f_{\rm cool}=1$. Finally, we can compute the mass accreted onto each planet at a given time step as

\begin{equation}\label{Macc}
M_{\rm acc}=\dot{M}_r \Delta t f_{\rm cool} f_{\rm hydro} {\rm min}(\frac{R_h}{H};1).
\end{equation}

We checked that with this new recipe to include planets in our simulations, our code still conserves mass and angular momentum throughout all the simulations.

 
\subsection{Model of the Sun's evolution}\label{sun}

In our fiducial model, we start the release of gas in the asteroid belt at 5 Myr, which is a good estimate for the end of the proto-planetary disk phase. It is confirmed by observations showing that aqueous alteration may have begun between $<$1.8 and 6.8 Myr after the formation of the CAI \citep{2023NatAs...7..309M,2024E&PSL.62718559M}, which can only happen when the snowline moves back outwards from $\sim$2 au in the proto-planetary disk to $>3$ au in the debris disk phase. We will discuss later what a change in this starting time would imply to our results.

\begin{figure}
   \centering
   \includegraphics[width=9.cm]{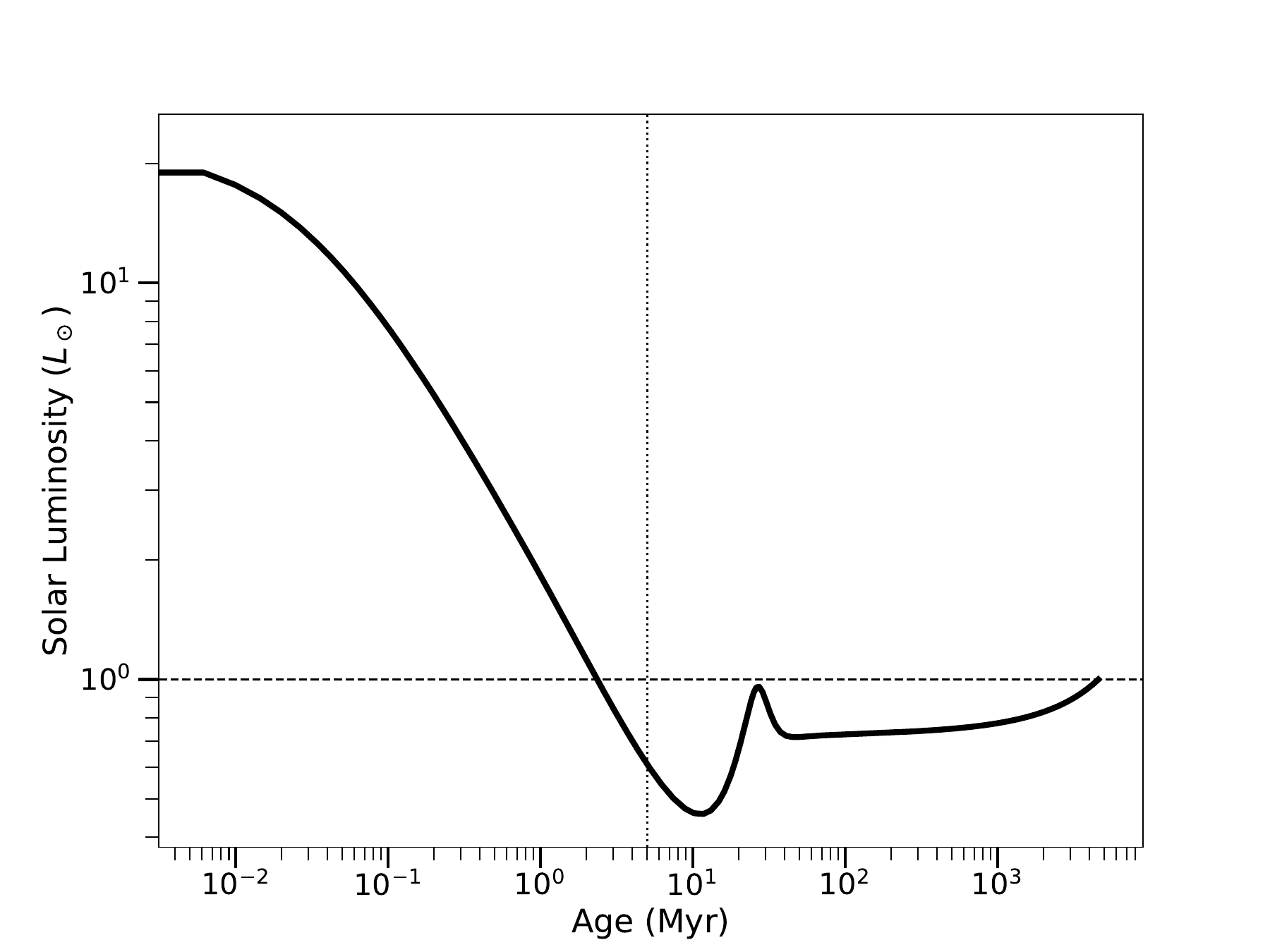}
   \caption{\label{figsun} Solar luminosity (in $L_\odot$) as a function of time in Myr (see details of the solar model in Sect. \ref{sun}). The horizontal dashed line is showing the current solar luminosity. Our gas release model starts at 5 Myr (vertical dotted line) when the proto-planetary disk dissipates and the sun's luminosity is around 0.6 $L_\odot$.}
\end{figure}

Figure~\ref{figsun} shows the output of a state-of-the-art code, \cesamxx \citep{1997A&AS..124..597M,2008Ap&SS.316...61M,2013A&A...549A..74M}, used to model the Sun and stellar evolution in general. \cesamxx  is used to simulate the Sun's evolution from its youth to its current luminosity as it is described in more detail in appendix \ref{appsun}. In Fig.~\ref{figsun} showing the Sun's luminosity as a function of time, we can see that initially the luminosity decreases because the Sun keeps on collapsing, becomes denser and cools. The Sun is initially completely convective with a large radius, but a radiative zone appears around 1 Myr. Between 20 and 90 Myr, an additional transient convective zone appears next to the core, as it is now sufficiently hot and dense to initiate the first stage of the CNO cycle (nuclear fusion converting hydrogen into helium via nuclear reactions involving carbon, nitrogen and oxygen), i.e. the conversion of $^{12}$C into $^{13}$N, while releasing a large excess of energy. This sudden excess of energy destabilizes the medium and leads to the formation of this internal convective zone next to the core. The excess energy is then released from the surface, leading to a luminosity surge that is clearly observed between 20 and 40 Myr. The second stage of the CNO cycle (conversion of $^{13}$N to $^{13}$C) requires much higher temperatures, so that the CNO cycle remains stuck at the first stage. Once all the $^{12}$C has been converted, this energy source runs out, the inner convective zone disappears, and luminosity drops sharply. 

But in the meantime, the temperature has risen sufficiently for the proton-proton chain (another set of nuclear fusion reactions by which hydrogen can be converted into helium) to get underway. There is a lot more fuel available, but the reaction is much weaker than the CNO cycle and does not generate enough energy to destabilize the medium, so luminosity grows slowly over time to reach the current solar luminosity (dashed line in Fig.~\ref{figsun}).

\section{Results}

\subsection{Evolution of the gas disk}\label{evolgas}

Gas is injected into the disk as it sublimates from the initially icy bodies in the asteroid belt. The $\dot{M}$ rate at which it is injected (integrated over all sizes and radii so that it is equal to $1/\delta r \int_{2.3 {\rm au}}^{3.3 {\rm au}} \dot{\Sigma}_{\rm io}(r) \mu_w m_p dr$) depends mainly on the initial mass of the belt once we have fixed the solar luminosity, the location of the icy belt and its size distribution. We run our models for 1 Gyr using the two scenarios presented earlier, where we start with the current asteroid belt mass $\sim 4 \times 10^{-4}$ M$_\oplus$ (scenario 1), and with a mass of 0.1 M$_\oplus$ that gets depleted to current asteroid belt level after 50 Myr (scenario 2). 

\begin{figure}
   \centering
   \includegraphics[width=9.cm]{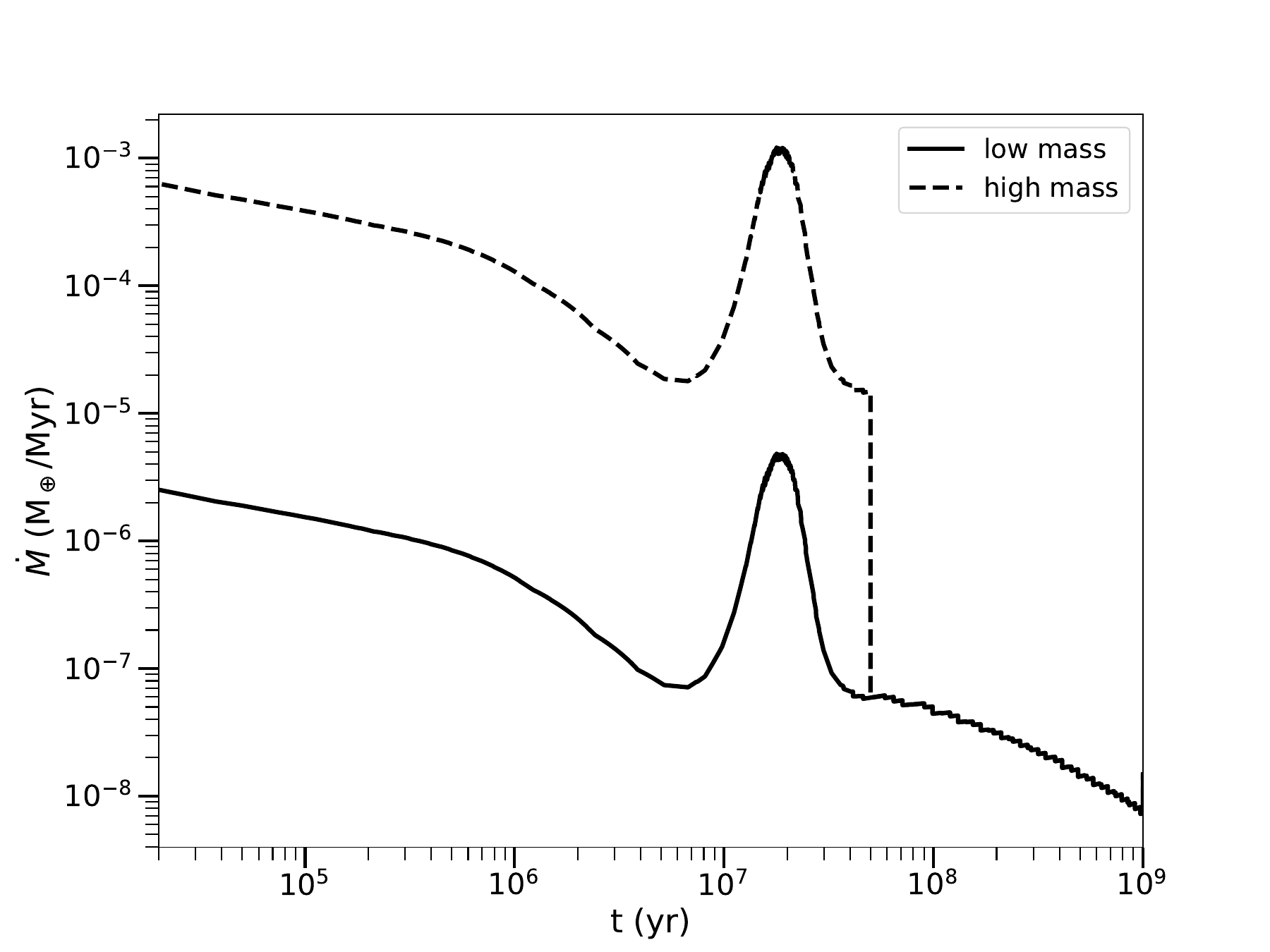}
   \caption{\label{figmdot} Injection rate of gas $\dot{M}$ from the asteroid belt into the gas disk as a function of time (since the start of sublimation, i.e. when the Sun was $\sim 5$ Myr) for scenario 1 (low mass) and scenario 2 (high mass).}
\end{figure}

In Fig.~\ref{figmdot}, we show the rate of injection of gas into the belt for both scenarios\footnote{Note that our simulations start when the proto-planetary disk is no longer present, i.e. after $\sim 5$ Myr and thus one needs to add 5 Myr to the time shown on the model's plots to retrieve the age of the Sun.}. We can see that the rate of injection into the belt is high at the beginning (but for a short period, beware of the logscale in time), as this is when the greatest mass is available for sublimation. Very quickly, the smallest bodies become completely rocky, and the ice sublimation can then only happen for the largest bodies. A peak appears around 20 Myr as the Sun's luminosity increases rapidly, leading to an increase in the temperature $T$ of solid bodies $\propto L_\star^{1/4}$ and an increase in $\dot{M}$, which scales as $\exp(-a_2/T)$ (see Eq.~\ref{Zeq}). The increase in $\dot{M}$ is a factor of $\sim$100 compared to the values before and after the surge, over a considerable period of time (tens of Myr) and it is in fact the period when most of the gas mass is released from the asteroids. The injection rates of the high (dashed line) and low mass (solid line) scenarios are a factor 250 apart, which corresponds to the ratio of the initial masses used in these two scenarios. $\dot{M}$ can reach values of the order of $10^{-3}$ M$_\oplus$/Myr in scenario 2 during the surge, which is also a typical value in colder exo-Kuiper belt releasing CO \citep[e.g.][]{2020MNRAS.497.2811K} and could have been even higher in the initially massive Kuiper belt beyond Neptune \citep{2021A&A...653L..11K}.

\begin{figure}
   \centering
   \includegraphics[width=9.cm]{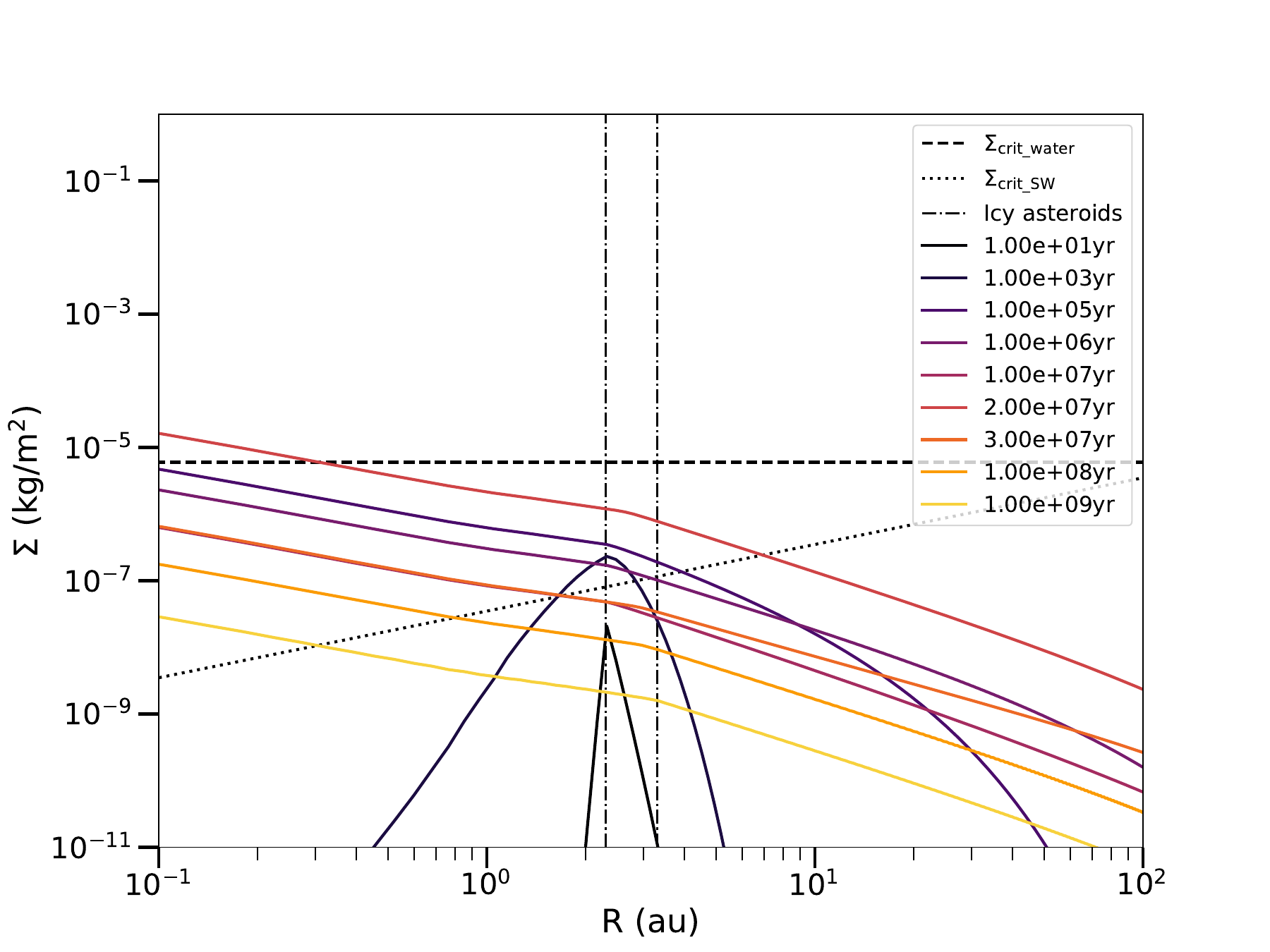}
      \includegraphics[width=9.cm]{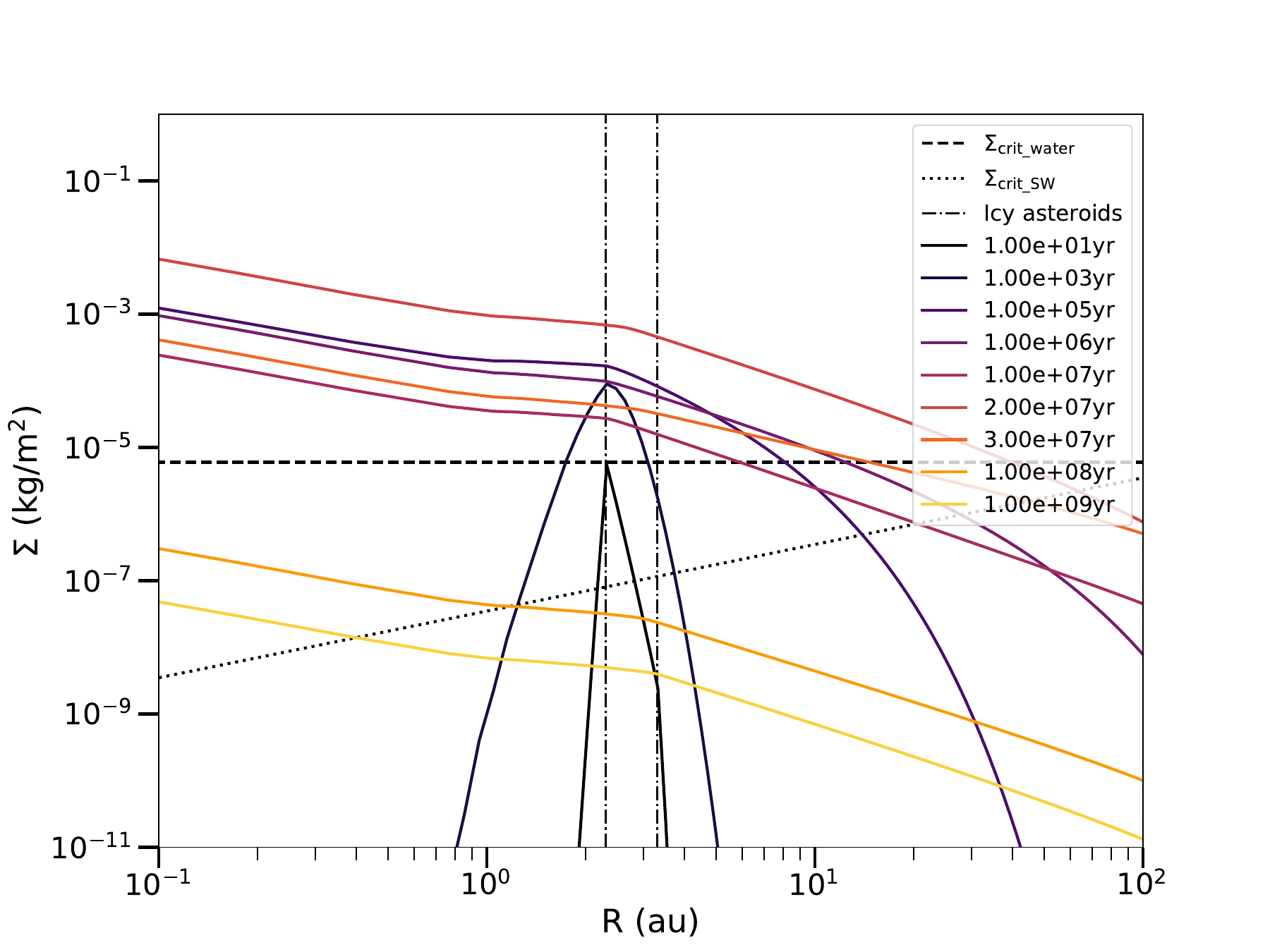}
   \caption{\label{figsigma} Evolution of the surface density profile as a function of time for scenario 1 (top) and scenario 2 (bottom). The different colours show the temporal evolution as indicated in the legend with brighter colours indicating later times. The dashed line is the density above which water is self-protected against dissociation and the dotted line is an upper limit of the critical density where gas will start to be blown away by the solar wind and create an outflow. The dash-dotted line shows the location of icy asteroids that release water in the belt.}
\end{figure}

The released gas will spread over time due to viscous evolution. In our fiducial model, we start with an $\alpha$ value of $10^{-2}$ but we later explore  how a change in this parameter could affect the results. In Fig.~\ref{figsigma}, we show the viscous evolution of the gas disk from $t$ close to zero, where we can notice that initially the gas is injected between 2.3 and 3.3 au, i.e. at the location of the initially icy asteroids, and that as time increases gas spreads inwards and outwards due to viscosity (see \citealt{2016MNRAS.461.1614K} and \citealt{2024MNRAS.530.1766C} to explore what mechanisms may explain the viscosity). The disk never reaches a steady state, as $\dot{M}$ varies over time and the viscosity varies with temperature (which depends on the varying solar luminosity). By observing the evolution of the radial profile of the surface density at different times, we can see that gas first accumulates in the vapour producing 2.3-3.3 au region and the disk then progressively spreads radially. Up to $\sim 10$ Myr, its surface density decreases because of the combination of the spread and the decrease of $\dot{M}$. Between 10 and 25 Myr, there is an upturn in the gas surface density because of the sudden increase of $\dot{M}$ (Fig.~\ref{figmdot}). Then, after 25 Myr, $\dot{M}$ decreases again and so does the disk density.

Viscous evolution takes place on time scales $t_{\rm visc} \sim r^2/\nu \sim 300/\alpha$ years in the asteroid belt. This means that for $\alpha=10^{-2}$ the gas injected into the main belt will take around 30,000 years to spread in significant quantities into the inner region close to the Sun. This can be thought of as the time needed to fully react to changes in $\dot{M}$, or to erase the memory of a given input rate. If $\alpha$ increases (decreases) by a factor of 10, $\dot{M}$ will remain the same\footnote{Section \ref{accsec} is dedicated to the accretion of gas onto planets and we note that $\alpha$ has little influence on gas mass accreted onto planets. If we consider the amount of mass that radially crosses a planet and average over timescales longer than the viscous timescale, we would end up with the same total mass accreted onto the planets and we  evaluate this later in the paper.} (as it is mostly fixed by the asteroid belt mass) but the viscous evolution timescale will become smaller (greater) by a factor of 10, thus lowering (increasing) $\Sigma$ by a factor of 10. 

Estimating the surface density in the gas disk is of crucial importance because two essential processes depend on its value. If the density is high enough then the gas disk can become optically thick to stellar light, effectively shielding the water molecules from photo-dissociation. The critical density at which the shielding occurs can be estimated to be $\Sigma_{\rm crit\_water}=6 \times 10^{-6}$ kg/m$^2$ (see Sect. \ref{theeq}).
On the contrary, for very low gas densities, water molecules can become affected by stellar wind \citep[see][]{2023A&A...669A.116K}. We estimate the density value below which this can happen with the formula 

\begin{equation}
\Sigma_{\rm crit\_SW}=\frac{2 H \mu m_p}{\Delta R \sigma_{\rm col}}=7 \times 10^{-8} \, {\rm kg/m^2} \left(\frac{r}{2{\rm au}}\right)\left(\frac{\Delta r}{1.5 {\rm au}}\right )^{-1},
\end{equation}

\noindent where we assumed $T_g=197$ K (blackbody temperature) at 2 au and a scaling of $r^{-0.5}$ (it is represented as a dotted line in Fig.~\ref{figsigma}). For the collisional cross-section, $\sigma_{\rm col}$, we took a polarisability of $\alpha_{\rm H_2O}=1.501$\AA\, \citep{1997CP....223...59O} and calculated the typical radius of the particle modeled by an imaginary hard sphere (Van der Waals radius) given by $(3\alpha_X/(4\pi))^{1/3}$ so that we obtain $R_{\rm H_2O}=0.71$\AA\, or $\sigma_{\rm col}=1.6 \times 10^{-20}$ m$^2$. We note that the typical radii for H and O are 0.54 and 0.58\AA, respectively, which would lead to similar cross-sections. This gives an upper limit for the critical density at which the solar wind can start pushing the gas away onto radial orbits, as taking into account collisions between protons at high velocities ($>100$ km/s) and H$_2$O would lead to a higher cross-section due to the permanent dipole of water\footnote{Indeed, the elastic cross-section between protons and water vapour at 1keV is $\sim 10^{-19}$ m$^2$ according to \citet{2012NIMPB.273...98C} and extrapolating to 100 eV, one finds $\sim 3 \times 10^{-19}$ m$^2$, which would lower the stellar wind critical density by a factor $\sim 20$ for H$_2$O. But this value is uncertain and may not be that useful in our setup given that water will have photo-dissociated when it reaches low enough densities to be blown out and we therefore do not use it in our study.}.

We could naively assume that as scenario 2 starts with a mass that is 250 times greater, the gas disk would then evolve with a surface density that is 250 times greater, but this is not the case. Indeed, the mean molecular weight of the atomic gas ($\mu=6$ because the gas disk is made up of atomic H and O rather than the H$_2$O molecule) is smaller in scenario 1 by a factor 3 compared to scenario 2, thus implying a greater viscosity (by a factor 3) and a lower viscous timescale by the same factor. It is similar to having a greater value of $\alpha$ so that the disk spreads faster and less gas can accumulate in scenario 1, compared to a gas disk with $\mu=18$ (scenario 2); thus  the total surface density would drop and lead to an increase in the 250 factor.
 
In the case of high mass (scenario 2), the surface density is greater than that of the water shielding critical density, and the gas remains mainly in the form of water molecules. In the low-mass case, water photo-dissociates into H+OH, then OH dissociates into O+H and we obtain a disk composed of atomic gas with an average molecular weight of 6. However, we note that for low $\alpha$ values ($<10^{-3}$), because the gas surface density gets higher, water can start shielding even in the low mass case scenario (scenario 1). For both scenarios, the gas density in the main belt may eventually (after typically a few 100 Myrs) becomes low enough for the solar wind to begin blowing the gas away (depending on the exact collisional cross-section between high velocity protons and H or O), creating a weak atomic gas wind that outflows rather than a viscously spreading disk, similar to the predictions for the current Kuiper belt \citep{2021A&A...653L..11K}.

\begin{figure}
   \centering
   \includegraphics[width=9.cm]{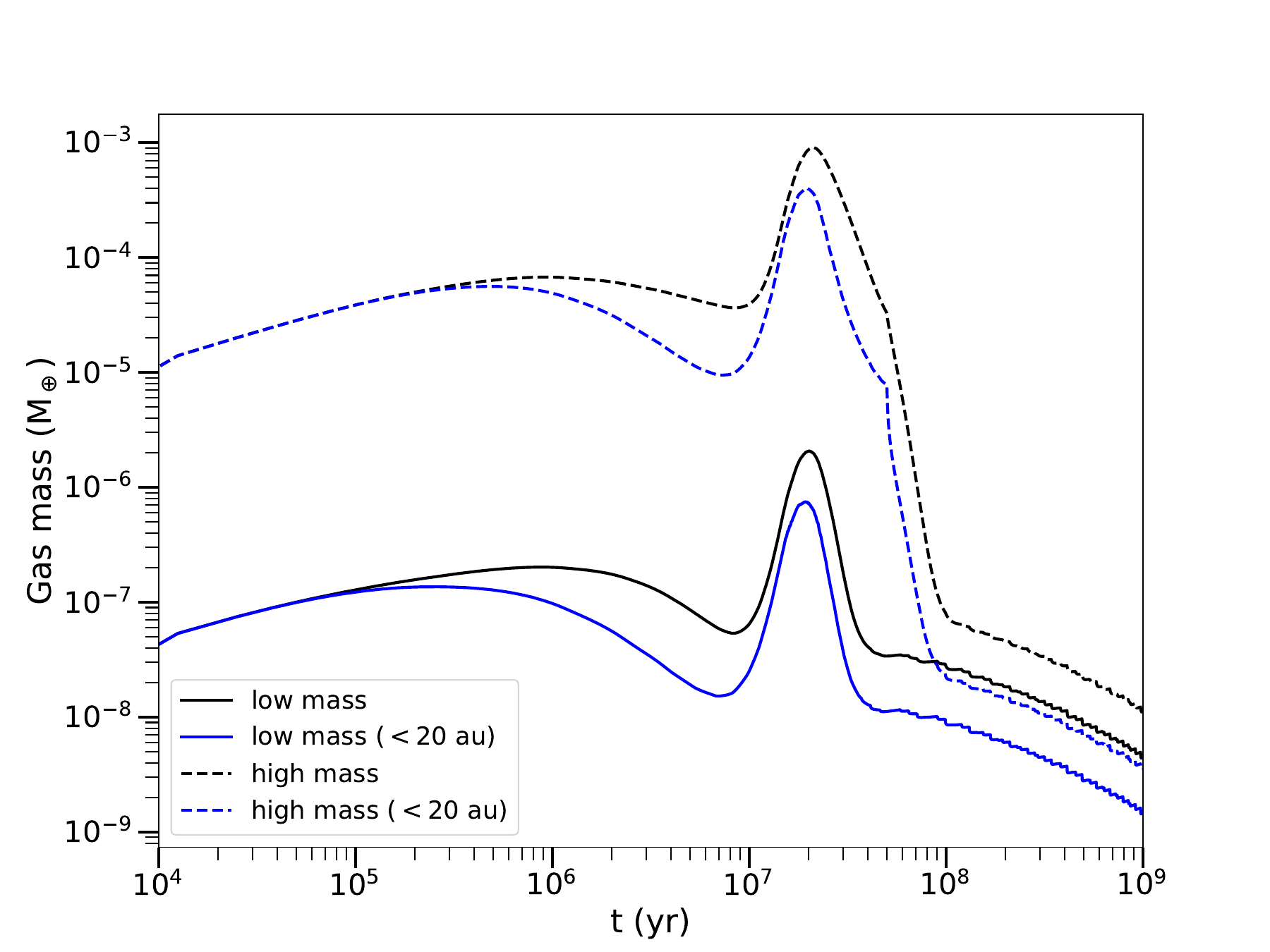}
   \caption{\label{figgasmass} Total gas mass (in M$_\oplus$) in the disk for scenarios 1 (solid) and 2 (dashed) as a function of time (in yr). The black lines are for masses including gas up to 400 au and the blue lines for up to 20 au.}
\end{figure}

In Figure~\ref{figgasmass}, we show the temporal evolution of the total gas mass in the disk for the low (solid) and high (dashed) mass scenarios (corresponding to scenarios 1 and 2, respectively). The mass located within 20 au is shown in blue and up to 400 au in black. The total gas mass is highest during the luminosity surge (at around 20 Myr) and can reach $2 \times 10^{-6}$ M$_\oplus$ for scenario 1 and $10^{-3}$ M$_\oplus$ for scenario 2. We note that the difference in gas mass between the two scenarios is not exactly 250 for the same reasons as the surface density (i.e. different $\mu$ in the two scenarios).

\subsection{Evolution of the ice content}

 \begin{figure}
   \centering
   \includegraphics[width=9.cm]{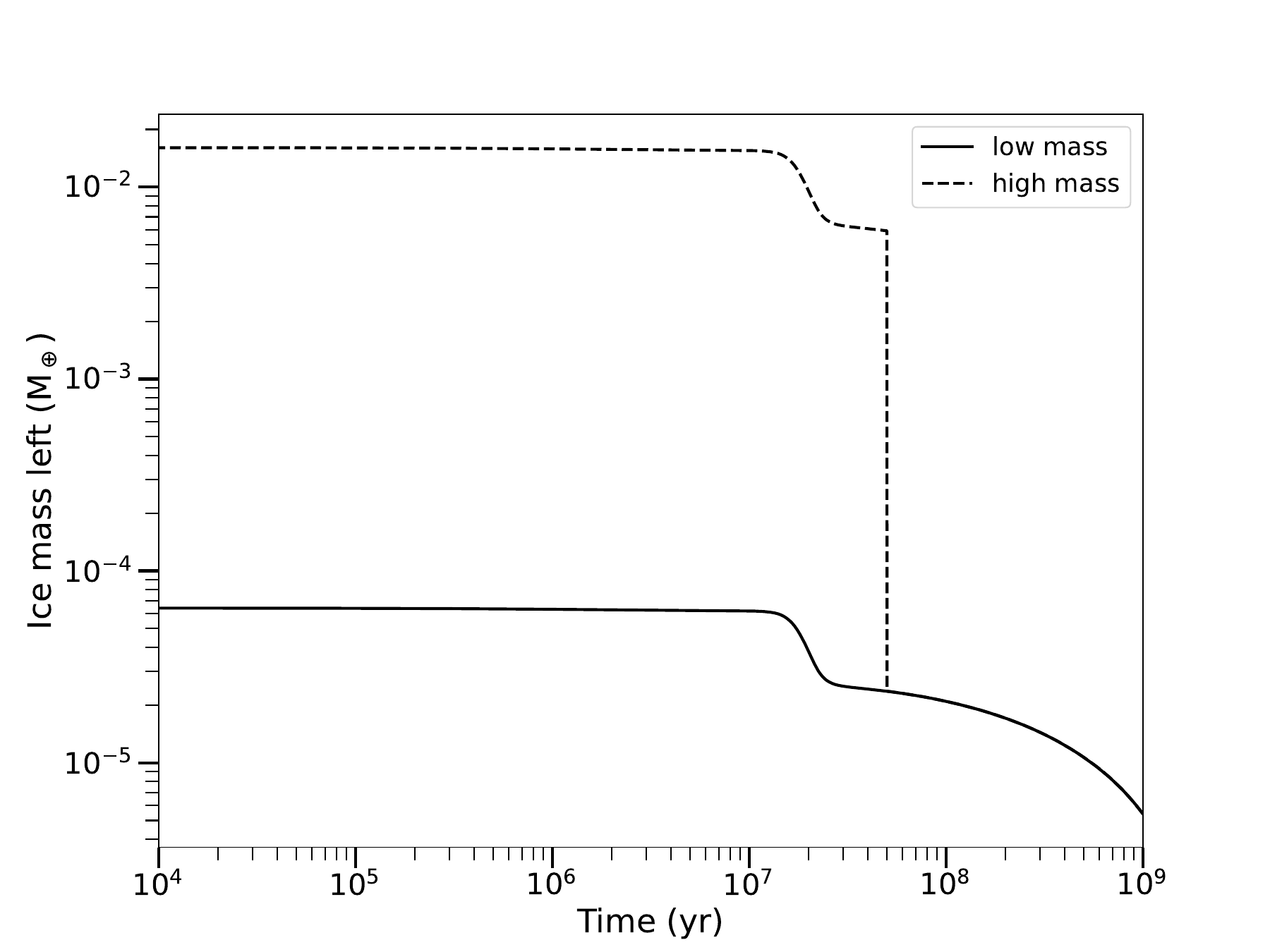}
   \caption{\label{figicetime} Total ice mass left on solid bodies as a function of time for the scenario 1 (solid) and scenario 2 (dashed). The depletion at 50 Myr for the high mass case (scenario 2) mimics an early instability that can deplete the asteroid belt.}
\end{figure}

It is also instructive to examine the temporal evolution of ice left in solid bodies as a function of time and size. Fig.~\ref{figicetime} shows that the total ice mass starts from $\sim 6.4 \times 10^{-5}$ M$_\oplus$ (low mass, solid line) and $1.6 \times 10^{-2}$ M$_\oplus$ (high mass, dashed line) for the scenarios 1 and 2, respectively. The qualitative evolution is then very similar in both cases with an almost constant level for the first 10 Myrs, followed by a sharp decrease during the luminosity surge around 20 Myr after the start of sublimation, and then a slow decrease up  until 1 Gyr where only $\sim$ 8\% of the initial mass remains for scenario 1. The sharp decrease at 50 Myr for the high mass case (scenario 2) is artificial as this is the moment where we choose to deplete the belt to mimic, e.g. an early instability, and the final mass is then $\sim$0.03\% of the starting mass. 

\begin{figure}
   \centering
   \includegraphics[width=9.cm]{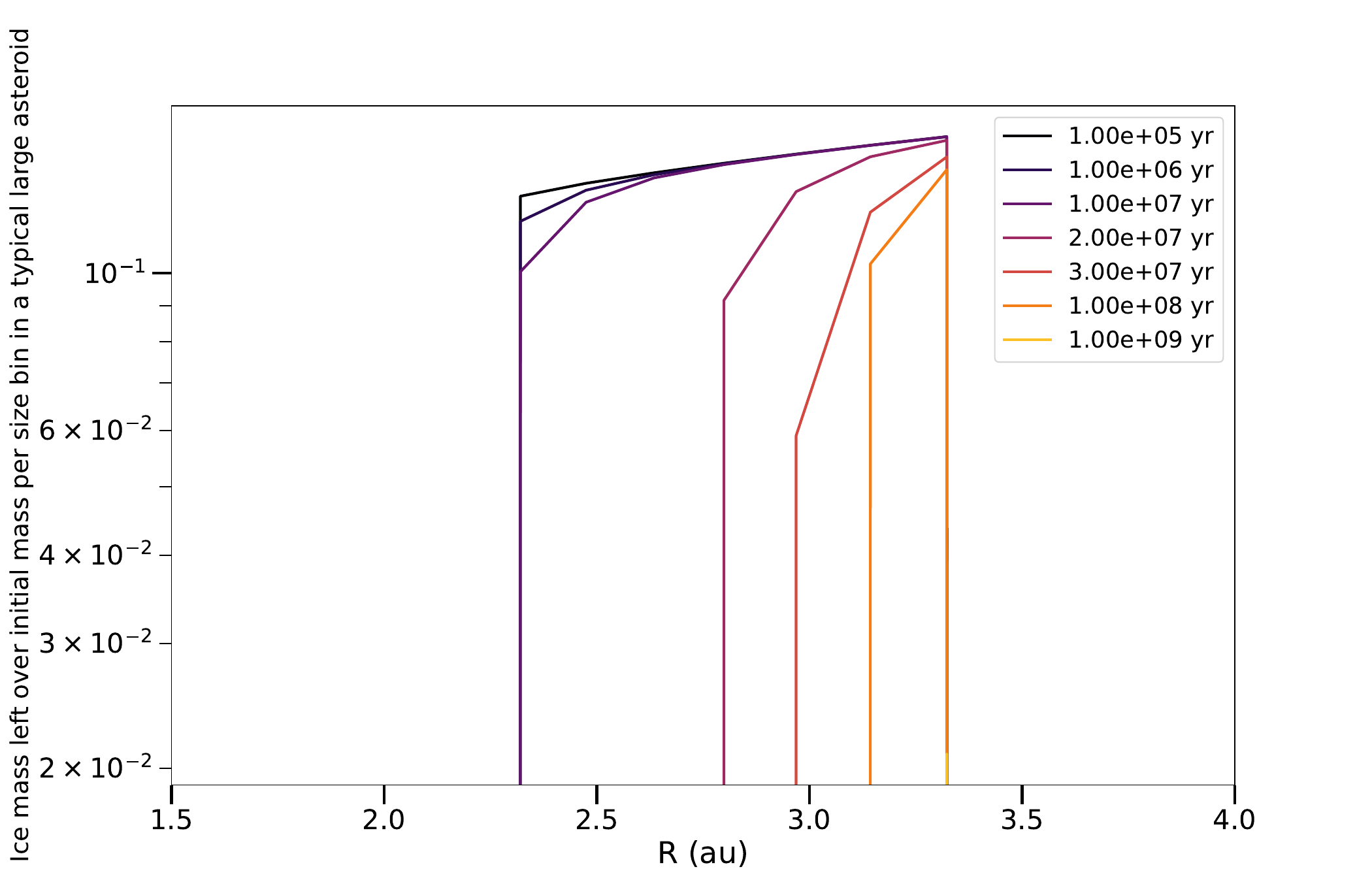}
   \caption{\label{figlargest} Ice mass left over total initial mass in the different radial bins in the main belt for a large 100 km body as a function of time (brighter colours are for longer times) for the scenario 1. Most of the mass gets depleted from the inner region after the surge in luminosity at $\sim$ 20 Myr. There is no ice left on the 100 km body after 1 Gyr evolution.}
\end{figure}

\begin{figure}
   \centering
   \includegraphics[width=9.cm]{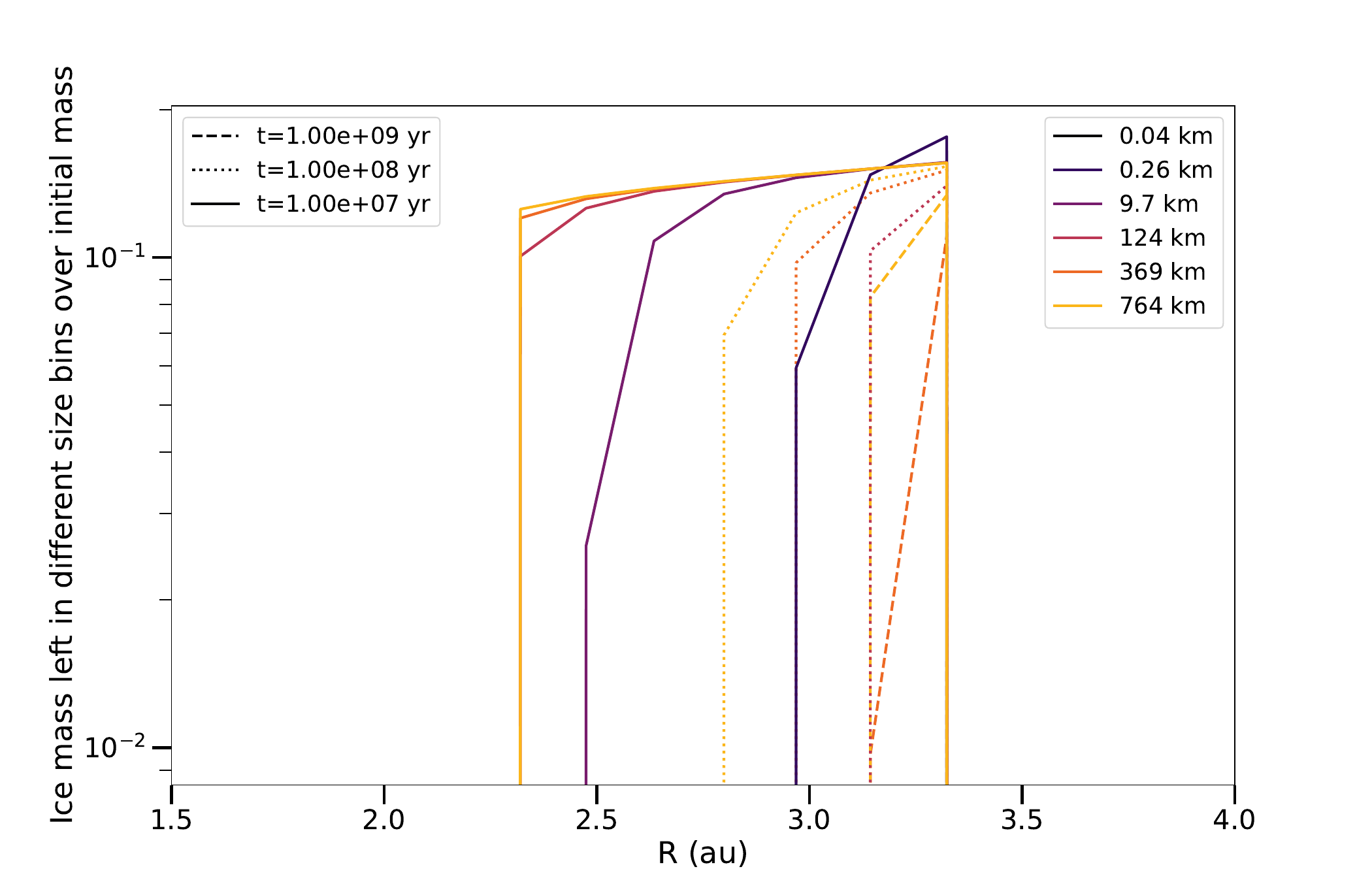}
   \caption{\label{figdiff} Ice mass left over total initial mass in different size bins (brighter colours are for larger objects) after 10 Myr (solid), 100 Myr (dotted) and 1Gyr (dashed) for the scenario 1. The smallest size bin ($\sim 40$ m, black colour) never shows up on the plot as it gets depleted before 10 Myr.}
\end{figure}

Figures~\ref{figlargest} and ~\ref{figdiff} show in more detail the evolution of the ice content as a function of radial distance and solid sizes. For the sake clarity, we only present the results for scenario 1, the results for scenario 2  being qualitatively very similar and essentially offset by a scaling factor of 250. Let us first focus on a large size bin (because small solid bodies lose ice too rapidly) and follow the ice depletion as a function of time. Fig.~\ref{figlargest} shows the remaining ice mass at different distances in the main belt for a $\sim 100$ km body as a function of time. We can indeed verify that for such large bodies, the inner belt gets depleted inside out and becomes devoid of ice after the surge in luminosity at $\sim 20$ Myr. After 100 Myr of evolution, only asteroids (for a size $\sim 100$ km) closest to 3.3 au remain icy but they eventually lose all their ice after 1 Gyr of evolution. This means that most primordial ice should indeed have disappeared in the current asteroid belt as we confirm with the next figure showing the evolution of even larger bodies. Fig.~\ref{figdiff} shows the remaining ice mass for different size bins (colour) and different times (line style), demonstrating that indeed the smallest objects are being depleted rapidly and that ice only remains in the outer parts of the main belt on the largest objects $\gtrsim 100$ km for more than 100 Myr evolution and only bodies close to 3.3 au and greater than $\sim$300 km still have ices after 1 Gyr of evolution.

\subsection{Accretion onto planets}\label{accsec}

\begin{figure}
   \centering
   \includegraphics[width=9.cm]{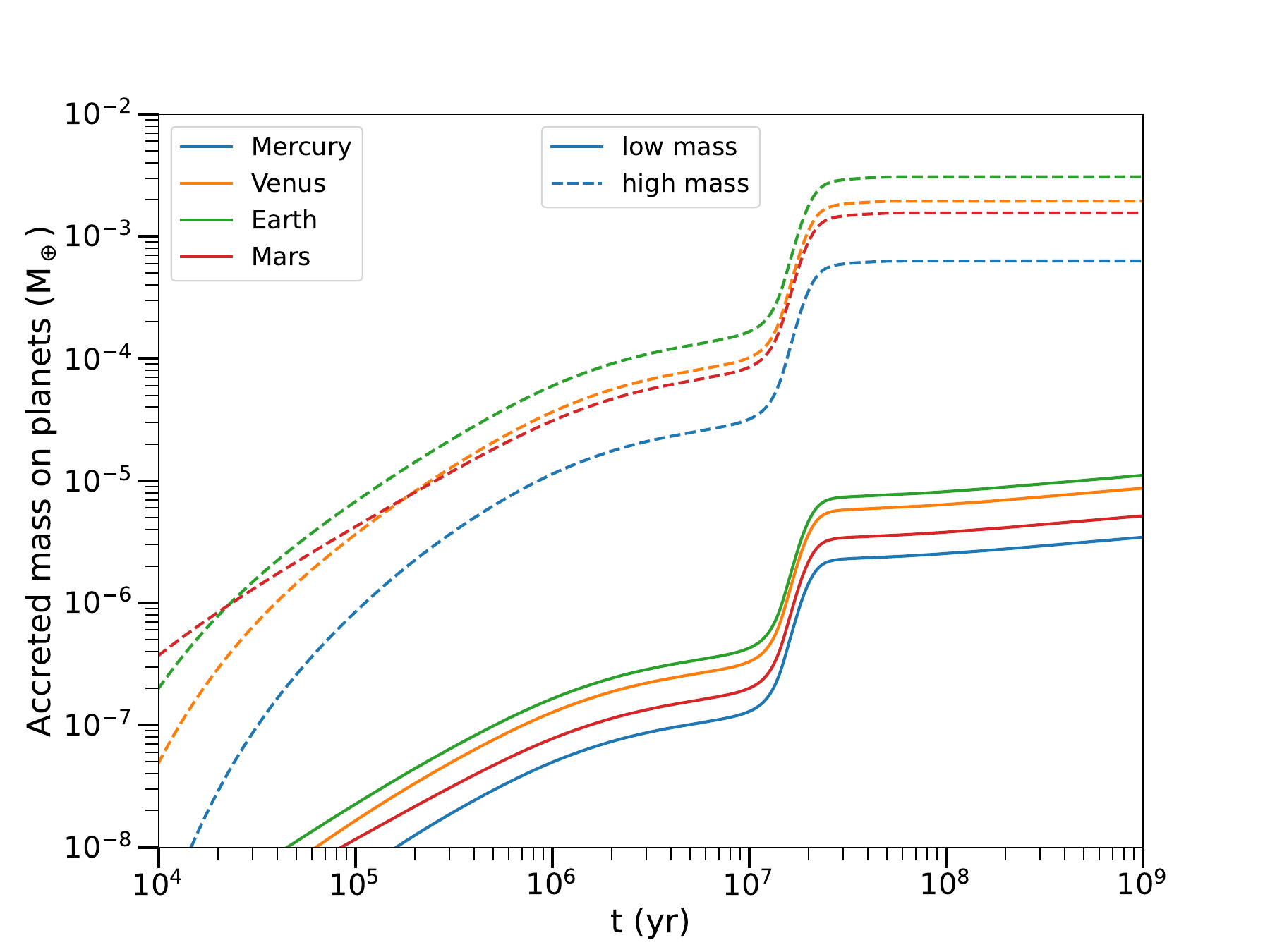}
   \caption{\label{figaccr} Cumulative mass accreted onto the different planets (Mercury, Venus, Earth, Mars - in blue, orange, green, red, respectively) as a function of time for the scenario 1 (solid) and 2 (dashed). We see that indeed most gas gets accreted during the surge in luminosity around 20 Myr.}
\end{figure}

We now focus on the mass accreted onto the 4 terrestrial planets Mercury (blue), Venus (orange), the Earth (green), Mars (red), and show its temporal evolution in Fig.~\ref{figaccr} for the low (solid) and high (dashed) mass scenarios (corresponding to scenarios 1 and 2, respectively). One notices that most of the water gets accreted around 15-25 Myr after the start of the simulation, which is when the surge in the luminosity of the Sun happens. Interestingly, the amount of water accreted on the different planets is not radically different in this mechanism (up to a factor 3 difference). In fact, it mainly depends on the radial flux $\dot{M}_r$ through each cell, which is relatively constant as a function of distance from the Sun, once the disk has had time to spread viscously. Note that Mercury and Mars will naturally accrete less gas, due to their smaller $R_h/H$ values close to 0.2 in our model while the Earth and Venus have values closer to 0.5. 

The Earth is the biggest accretor even though its $R_h/H$ value is similar to that of Venus (and equal to 0.49; see Sect. \ref{modpla}) because it is located closer to the asteroid belt and it gets served first thus reducing the gas quantity available for Venus. After 1 Gyr evolution, the total mass accreted onto all planets is $3 \times 10^{-5}$ M$_\oplus$ for scenario 1 and $7 \times 10^{-3}$ M$_\oplus$ for scenario 2 (approximately 250 times more), which represents $\sim 50$\% of the ice mass released after 1 Gyr. The rest of the gas is made up of two components, one is the gas disk mass as shown in Fig.~\ref{figgasmass} and the other the gas lost to the star at the inner edge and at the outer border of the simulation. 

In Table~\ref{tabaccr}, we list the total gas masses accreted by the different planets after 1 Gyr evolution for both scenarios. To get a comparison point, we note that the total water mass on the Earth is estimated to be around 1-10 oceans of water, including water within the mantle, or between $2.3 \times 10^{-4}$ to $2.3 \times 10^{-3}$ M$_\oplus$. The high mass case (scenario 2) result could then explain the presence of the majority of water on the Earth while the low mass case scenario (scenario 1) would only explain a portion of it. We will discuss these crucial results concerning the Earth and other planets in more detail in Sect. \ref{delivery}.


\begin{table}[h!]
  \begin{center}
    \caption{Total water mass (in M$_\oplus$) accreted as gas onto the different planets after 1 Gyr evolution.}
    \label{tabaccr}
    \begin{tabular}{c||c|c|c|c} 
     Scenario & \textbf{Mercury} & \textbf{Venus} & \textbf{Earth} & \textbf{Mars}\\
      \hline
      Low mass & $3\times 10^{-6}$ & $8 \times 10^{-6}$ & $10^{-5}$ & $5 \times 10^{-6}$\\
       \hline
      High mass & $6\times 10^{-4}$ & $2\times 10^{-3}$ & $3\times 10^{-3}$ & $1.5 \times 10^{-3}$\\
    \end{tabular}
  \end{center}
\end{table}

\section{Parameter exploration}

Thanks to the simplicity of the model, we do not need to run more numerical simulations and can explore the parameter space using some analytical reasonings.

\subsection{Change in the $\alpha$-viscosity parameter}\label{alphavar}

In debris disks, $\alpha$ is not very well constrained as it depends on molecular viscosity or complex hydro instabilities (e.g. the VSI) or even MHD instabilities (e.g. the MRI) as explained in more detail in \citet{2016MNRAS.461.1614K,2024MNRAS.530.1766C}. A change in $\alpha$ will have effects on the gas disk. Indeed, the gas disk viscosity is proportional to $\alpha$ and the viscous timescale scales as $1/\alpha$. Decreasing $\alpha$ will thus lead to an increase of the residence time of gas in the disk (before it gets accreted onto the star) and the total gas surface density (given enough time) will increase by a factor $\alpha$. However, the input rate of gas into the belt $\dot{M}$ does not change nor is the radial mass flux going through each cell $\dot{M}_r$\footnote{because $\dot{M}_r=-2 \pi r v_r \Sigma$ and $\Sigma$ increases when $\alpha$ goes down while $v_r$ decreases by the same quantity, thus canceling their effects.} and our results concerning the total mass accreted onto planets will not be affected. 

To check the last point more firmly, we ran one simulation with a lower $\alpha$ value. We notice that the only difference is the time for the gas to reach planets and start accreting because it takes longer to spread as expected by the greater viscous timescale $t_{\rm visc} \sim 300/\alpha$ years at the belt location (see Sect. \ref{evolgas}). A side effect is that it means that if $\alpha$ becomes smaller than $10^{-5}$ so that $t_{\rm visc}>30$ Myr, the total accreted mass on the planets would not change but it may take longer than 30 Myr to get it accreted onto the planets. Thus, it may change some of our conclusions on the period where most accretion happens, which would become decorraleted to the surge in luminosity and would happen at, e.g. $\sim$300 Myr for $\alpha=10^{-6}$. However, $\alpha$ values lower than $10^{-5}$ may not be very realistic because in low density gas disks such as those studied here, the molecular viscosity can become important leading to $\alpha$ values greater than $10^{-5}$ \citep{2024MNRAS.530.1766C}.

\subsection{Change in gas temperature}

In our model, we assumed a gas temperature, $T_g$, which is the same as that of the solid bodies. It means that we assumed that $T_g$ scales as $r^{-1/2}$, which may not be correct and that the absolute temperature is given by the black body approximation, which might also be false. For instance, one could imagine that in the less dense disks (not shielded and dominated by atoms), the gas temperature could go up due to its inability to cool. In fact, it is very complicated to calculate the temperature self-consistently because many heating and cooling mechanisms can intervene \citep[e.g. see][]{2016MNRAS.461..845K,2017MNRAS.469..521K}, some of which depending on the unknown composition or unknown quantities. In a simplified analytical model in \citet{2017MNRAS.469..521K}, we show that accounting for carbon photo-ionisation and recombination with electrons, the temperature does not increase exponentially but remains rather close to a black body temperature. However, there could be some large variations and therefore, we  simply analyze the impact that a change in the absolute temperature and profile would have without trying to further quantify the gas temperature. First, a change in the absolute temperature of the gas will have an effect on the viscosity because $\nu=\alpha c_s^2/\Omega \propto T_g$ and the end effect is similar to a change in $\alpha$ for the surface density (see previous paragraph). For example, an increase in $T_g$ by a factor of 3 would lead to a decrease in the surface density of the gas by a factor of 3. This will also change the scaleheight of the disk $H=c_s/\Omega \propto T_g^{1/2}$, which could affect the total mass accreted on the planets as it depends on ${\rm min}(\frac{R_h}{H};1)$ (see Eq.~\ref{Macc}). For example, an increase in $T_g$ by a factor of 3 would lead to a decrease in mass accreted onto the planets by a factor of 1.7 if it is in the regime where $\frac{R_h}{H}<1$. We also note that the gas surface density profile will be affected by a change in the temperature slope $\beta$. Indeed, the gas surface density profile at steady state is expected to vary as $\beta-x$ with $x=3/2$ interior to the belt and $x=2$ exterior to it \citep{2019MNRAS.489.3670K}. On the other hand, the radial mass flux at steady-state going through each cell $\dot{M}_r \propto r^{3/2-x}$ is independent of $\beta$ and thus this will not imply changes on the planet mass accretion rates.

\subsection{Change in initial ice mass}

The final results depend on the initial ice content as shown via our low and high mass cases (scenarios 1 and 2, respectively). For instance, an initial ice mass that is twice as large will lead to a release rate $\dot{M}$ that is twice higher and twice as much mass accreted on the different planets. In a way, our study shows that the initial ice mass could not have been much greater than that for scenario 2, i.e. $\sim 2 \times 10^{-2}$ M$_\oplus$ because then too much water gets accreted on, e.g. the Earth. This fixes an upper limit on the initial asteroid belt mass of $\sim 0.1 \times \frac{f_{\rm ice}}{0.2}$ M$_\oplus$, where $f_{\rm ice}$ is the fraction of ice-to-solid on initial asteroids. This is in agreement with models showing that the initial mass of the asteroid belt could have reached $\sim 0.5$ M$_\oplus$ \citep[e.g.][]{2000M&PS...35.1309M,2001Icar..153..338P} and is much smaller than the upper limit of 2 M$_\oplus$ \citep[e.g.][]{2007Icar..191..434O,2018ApJ...864...50D,2019AJ....157...38C} fixed using numerical simulations.

\subsection{Change in starting luminosity}

The moment in time when we start the luminosity evolution could be important. Here, we decided to start the gas release after 5 Myr of the Sun's birth when its luminosity is close to 0.6 $L_\odot$. Starting between 5 and 10 Myr would not change our results because most of the gas mass is released during the surge in luminosity at $20-30$ Myr. On the other hand, starting at 2 Myr when the luminosity is close to 1 $L_\odot$ would increase the initial solid body temperature by a factor $\sim 1.14$, which would boost the evaporation rate by a factor $\sim 2.7$ over the first few Myr. We note that this boost will not change our results by a large factor because most of the mass will still be released over the longer period of time ($> 10$ Myr, mind the logscale in Fig.~\ref{figsun}) when the luminosity surge happens.
 
\subsection{Change in albedo}

If we take the current Bond albedo of C-type asteroids that is roughly 0.02 \citep{1989aste.conf..557H,2022PSJ.....3...95V} instead of the 0.06 assumed, we would have a gas input rate that would increase by a factor 1.5 (see Eq.~\ref{Tbb}), thus speeding up the process. On the contrary, the initial albedo may have been even greater than the fiducial albedo assumed in this study. In this case, the temperature of solid bodies would have been smaller by a factor $(1-A)^{1/4}$, thus changing by a factor 1.04 if going from an albedo of 0.06 to 0.2. This is not a great factor but because of the exponential factor on the evaporation rate, the latter could be reduced by a factor $\sim 0.2$ and the evolution would then be slower overall. However, we note that the surface of active comets like 67P/CG are dominated by refractory material and that the detected exposed water ice represents only 0.1\% of the nucleus surface \citep{2023A&A...672A.136F} giving an albedo very close to 0.06 \citep{2022MNRAS.516.5125D}. Primordial icy asteroids could actually be dark, similar to active comets, and the albedo used in our model may be reasonable.

\subsection{Change in accretion efficiency}\label{eff}

In the fiducial model, we assumed that the accretion efficiency due to cooling is maximum with $f_{\rm cool}=1$. This is based on the model by \citet{2020NatAs...4..769K} where we have computed at each time step the theoretical accretion rate $\dot{M}_t$ due to cooling and compared to the actual potential accretion rate $\dot{M}_r$. Using a dust-free atmospheric composition \citep[e.g.][]{2014ApJS..214...25F,2015ApJ...811...41L}, we find that $\dot{M}_t$ is always greater than $\dot{M}_r$ for all planets considered, even during the luminosity surge when $\dot{M}_r$ increases greatly. Therefore, the accretion rate onto the planets is controlled by $\dot{M}_r$ rather than by cooling. However, we note that the specifics of $\dot{M}_t$ depend on the early atmospheric mass and composition of the planets, which is very hard to predict. Even though it is expected from the analytical model of \citet{2020NatAs...4..769K} that $\dot{M}_t$ is rather large, a more complex numerical model of planetary accretion would be useful to consider such complex young systems, which leaves some uncertainties on the $f_{\rm cool}$ parameter.

We have also considered that the amount of mass that gets in the Hill sphere without being accreted is $f_{\rm hydro}=1/2$. This recipe has been used widely in the proto-planetary disk community with varying values of $f_{\rm hydro}$ and some refinements such as considering that more mass gets removed farther away in the Hill sphere \citep[e.g.][]{1999MNRAS.303..696K,2018A&A...617A..98R,2020A&A...643A.133B}. Those results have been solidified lately with the use of 3D hydro simulations showing that gas flow is complex and can, for instance, circulate back to the circumstellar disk or just circulate around the planet (e.g. in a circumplanetary disk) without leading to accretion \citep[e.g.][]{2019A&A...630A..82L,2019A&A...632A.118S,2022A&A...661A.142M}. Considering those 3D simulations, $f_{\rm hydro}=1/2$ seems to be on the low side but specific simulations with terrestrial planets in a debris-disk like configuration would be required to confirm that.

\section{Discussion}

\subsection{Delivery of water to planets and the Moon}\label{delivery}
\subsubsection{Mercury}
Some water ice hides in Mercury close to its poles in regions of permanent shadow \citep[e.g.][]{2011Icar..211...37H,2013Sci...339..296N}. A current total mass of water ice of $3.4 \times 10^{-12}-3.4 \times 10^{-10}$ M$_\oplus$ is inferred from the MESSENGER spacecraft data \citep{2013Sci...339..292L}. Thermal models show that water ice could be stable in cold traps near the poles in permanently shaded regions over geological timescales \citep[e.g.][]{1992Sci...258..643P} given the long-term stability of Mercury’s obliquity.

It has been suggested that external sources may be necessary to explain the water found on Mercury because of the plausible dryness of Mercury’s crust and mantle meaning that interior sources such as volcanic outgassing may prove difficult to release large quantities of water, even over billions of years \citep{1999Icar..137..197M}. Impacts by interplanetary dust particles, asteroids or comets may explain the low water content on Mercury but it is not very efficient due to Mercury’s low gravity, because most of vaporized debris after impacts have velocities higher than the escape velocity and are not retained by the planet \citep{1999Icar..137..197M}.

Let us quantify the fraction of the delivered water to Mercury ($3\times 10^{-6}$ M$_\oplus$ for scenario 1 and $6\times 10^{-4}$ M$_\oplus$ for scenario 2) that could survive at the poles over geological time scales with our gas delivery mechanism. According to \citet{2013Sci...339..292L}, the area of permanently shadowed regions at both poles amounts to $\sim5.6 \times 10^{10}$ m$^2$, which is roughly $7.5 \times 10^{-5}$ of the total surface area of Mercury ($\sim 7.5 \times 10^{13}$ m$^2$). As an order of magnitude and assuming that the water delivered via our mechanism arrives uniformally on the planet, we find that the water trapped in the permanently shadowed regions would be $2.3 \times 10^{-10}$ M$_\oplus$ for scenario 1 and $4.5 \times 10^{-8}$ M$_\oplus$ for scenario 2 if all the water is retained over Gyr timescales, which is higher than the currently observed mass of $3.4 \times 10^{-12}-3.4 \times 10^{-10}$ M$_\oplus$.

However, we note that some external factors could reduce the delivered water content such as the erosion by
micrometeoritic impact on exposed ice deposits \citep{1999Icar..137..197M}. The erosion rate is uncertain and depends on many factors. For instance, once the ice is covered with fine layers of dust, it can be protected from sputtering, micrometeorite impacts, and sublimation \citep{1999Icar..137..197M}. Given the uncertainties on this erosion rate, one needs a scenario where the delivery rate is high enough that it can explain the current water mass estimates to suggest a plausible way of delivering water to Mercury, which is what our mechanism provides.

\subsubsection{Venus}

 The present-day water atmospheric mass of Venus is $\sim 10^{-9}$ M$_\oplus$ \citep{2018SSRv..214...10M}, which corresponds to $\sim 5 \times 10^{-6}$ Earth ocean mass. This does not account for water that could be trapped in Venus' mantle though no estimates exist to quantify that amount \citep[see discussion in][]{2009E&PSL.286..503G}. Therefore, it appears that Venus is very dry today but it could have been wetter in the past. Indeed, the D/H ratio in Venus of $120\pm40$ the telluric ratio \citep{1991Sci...251..547D} is extremely high, which has been used abundantly in the literature to infer a large primordial content of water accounting for Jeans escape over several Gyr \citep[e.g.][]{1982Sci...216..630D}. 
 
However, more recent works put that large water content into question \citep[e.g.][]{1993Natur.363..428G}. Indeed, escape mechanisms are not very well known and they could have varied over time. For instance, Venus could have suffered hydrodynamic escape in its youth, Jeans escape around 100-500 Myr, and at later times pick-up ion escape \citep{2009E&PSL.286..503G}. Venus express mission even shows that water (or rather O and H) is still escaping today at a significant rate due to the solar wind \citep{2017SSRv..212.1453F}. On the other hand, some water may not be primordial because it could have been resupplied at a later stage, either via volcanism and/or exogenic sources such as comets in the most recent $\sim 10^9$ years of planetary evolution \citep{2013tucv.book.....B}. Hence, the large D/H ratio results from the complicated history of escape over very long Gyr timescales and of possible recent water sources, which are very hard to properly quantify.

Therefore, the original water content in Venus is highly unconstrained. However, given the great similarity in bulk densities and its close proximity with the Earth, it may seem reasonable that Venus and Earth started with similar water abundances, which is the best assumption we can make at present. Because the mechanism we suggest in this paper could deliver the right amount of water to the Earth, and that it would bring roughly an equal amount to Venus, we can state that our new delivery scenario would in that sense satisfy this assumption.

\subsubsection{Earth}

The total amount of water on the Earth is estimated to be between $2.3-23 \times 10^{-4}$ M$_\oplus$, including water within the mantle. Today, the escape rate of hydrogen from the Earth is very low \citep[$\sim 3$ kg/s,][]{2017aeil.book.....C}. It is limited by the low mixing ratio of water vapor ($\sim$1-5 ppm in the upper atmosphere), due to the so-called cold trap (i.e. cloud formation) at the tropopause. Therefore, it is not very likely that the total amount of water on the Earth, once accretion was over, has changed significantly over geological time scales due to the water cycle, the cold trap and hence the very low escape rate. However, during the Archean, the escape rate of hydrogen could have been $\sim$2 orders of magnitude higher and was controlled by the mixing ratio of CH$_4$. CH$_4$ is indeed photolysed in the upper atmosphere and was more abundant \citep[likely 100-1000 ppm,][]{2001Sci...293..839C} during the Archean than today. 


The current deuterium to hydrogen ratio (D/H) for ocean water is known very precisely to be $1.5576 \pm 0.0001 \times 10^{-4}$ with respect to the Vienna Standard Mean Ocean Water (VSMOW). It is still a matter of debate whether this value is representative of the bulk of Earth’s water because of some measurements of a lower D/H ratio in deep mantle materials that may be more primordial \citep{2015Sci...350..795H}. The D/H ratio is a mixture of all sources that could have contributed to Earth's reservoirs and is abundantly used to identify the sources of Earth's water. As already stated in Sections \ref{intro} and \ref{whywaterice}, it appears that the best match to the current ocean water D/H ratio is that of carbonaceous chondrites, in particular CI and CM types \citep{2012Sci...337..721A} suggesting that they were the principal source of Earth’s water.

In this paper, we find that $10^{-5}$ M$_\oplus$ (scenario 1, low mass) and $3\times 10^{-3}$ M$_\oplus$ (scenario 2, high mass) of water could have been accreted onto the Earth, mostly between 20 and 30 Myr after the Sun's birth, via our viscous delivery mechanism. Therefore, the high mass scenario (scenario 2) could deliver all of Earth's water and even in the low mass case (scenario 1) this disk-delivered component cannot be neglected ($\sim 4$\% of an Earth ocean). 

It is also important to note that most of the water content of asteroids is sublimated rapidly in our model and the D/H in the gas state is then expected to be the same as that on the solids. Hence, within our mechanism, the D/H should be the same as that of initially icy bodies in the main belt, i.e. mainly C-type asteroids, whose D/H is indeed close to that of Earth's water \citep{2012Sci...337..721A}. We note that our mechanism does not need impact to deliver water and it would be interesting to find ways to distinguish both the impact and viscous delivery mechanisms, which we attempt in Sects. \ref{sign} and \ref{obswat}.

\subsubsection{Mars}

Today, it is estimated that 34 m GEL\footnote{GEL stands for Global equivalent layer and $1.4 \times 10^{18}$ kg of water corresponds to $10$ m GEL on Mars.} (or $8.0 \times 10^{-7}$ M$_\oplus$) of water lies in the Martian polar-layered deposits and shallow ground ice \citep{2015GeoRL..42..726C}. It is a lower limit because some additional water ice could hide deeper underground, notably within hydrated minerals depending on assumptions on the thickness of the hydrated crust \citep{2012LPI....43.1539M}. 

As for the quantity of water on the early Mars, there are quite large uncertainties. If oceans were present at the surface then a GEL of 100 m to 1 km liquid water would have been required near the surface \citep[e.g.][]{1991Natur.352..589B,2001Icar..154...40C}. By measuring the current D/H ratio in different locations on Mars, \citet{2015Sci...348..218V} find that water has a representative D/H value enriched by a factor of $\sim 7$ with respect to VSMOW. This is indicative of great water loss and they estimate that early Mars could have had about 137 m GEL ($3.2 \times 10^{-6}$ M$_\oplus$), which is in agreement with subsurface observations with the MARSIS/Mars Express radar data \citep{2012GeoRL..39.2202M}. More recently, \citet{2021JGRE..12606351W} estimated the amount of water required to form hydrated minerals observed on Mars and found 70-860 m GEL, which corresponds to a lower limit of $\sim 2 \times 10^{-6}-2 \times 10^{-5}$ M$_\oplus$.

Our model predicts that up to $1.5 \times 10^{-3}$ M$_\oplus$ of water could be accreted onto Mars in scenario 2, which is roughly larger by two orders of magnitude to these lower limits of early quantities of water on Mars. Within the framework described in this paper, this could be due to several reasons: 1) either Mars' history differs from other planets concerning its dynamics (e.g. it moved significantly due to an instability), or 2) its water inventory is underestimated, or 3) the accretion efficiency may be lower than for larger planets (see discussion in Sect. \ref{eff}) leading to less water accreted, or 4) some unknown physics is missing in the model.

\subsubsection{Moon}

Water ice is now detected on the Moon but its origin is still strongly debated \citep{2018PNAS..115.8907L}. The quantity of water ice in the permanently shadowed regions on the Moon is very uncertain and estimated to be around $\sim 10^{-11}$ M$_\oplus$ if all cold-traps hide a $\sim$10-m-thick pure subsurface ice deposit \citep{2019NatGe..12..597R} but it could be around two orders of magnitude lower based on surface detections and the LCROSS impact results \citep{2010Sci...330..463C,2015Icar..255...58H,2017Icar..292...74F,2018PNAS..115.8907L}. The Moon's axis may have varied over time and rotated to its current axis $\sim 3$ Gyr ago \citep{2016Natur.531..480S}, which allowed for water ice to accumulate over billions of years. The permanently shadowed regions are expected to have slightly evolved over time with a current surface of $\sim 36,000$ km$^2$ for both the North and South poles taken poleward 80°\footnote{We do not consider shadowed regions close to the equator because they are often not cold enough to be cold traps that can keep ices because of terrain irradiance \citep{2023LPICo2806.1806S}}. 

Because the Moon is within the Hill radius of the Earth, we can use our estimated accretion rate for the Earth to derive an accretion rate for the Moon after a few corrections. Using our model (see Fig.~\ref{figaccr}), we can estimate that the Earth accretes of order a few $10^{-6}$ M$_\oplus$ after the moon-forming impact (i.e. after 30-100 Myr). The Moon being smaller than the Earth, we need to correct for the smaller Hill radius by a factor $(M_{\rm moon}/M_\oplus)^{1/3} \sim 0.23$ leading to $\sim 5 \times 10^{-7}$ M$_\oplus$ that may accrete on the Moon. Assuming that water is accreted uniformly on the Moon and that only the water in the shadowed regions survives, we calculate that the ratio of the shadowed area to the total surface area ($3.79 \times 10^{7}$ km$^2$) is equal to $9.5 \times 10^{-4}$ and therefore find that the surviving water mass is $\sim 5 \times 10^{-10}$ M$_\oplus$, which is very close to the current estimates of water ice mass on the Moon. 

In addition, we note that there is $\sim 5 \times 10^{-7}$ M$_\oplus$ of water ice that has not yet sublimated from large bodies after 1 Gyr (see Fig.~\ref{figicetime}), which is expected to turn into vapour in the next 3.6 Gyr evolution up to now, of which roughly 50\% will be accreted onto planets and 15\% onto Earth according to our previous results. Accounting for the differences in Hill radii between the Earth and the Moon, we find $\sim 2 \times 10^{-8}$ M$_\oplus$ could be accreted on the Moon and $\sim 2 \times 10^{-11}$ M$_\oplus$ could survive in the permanently shadowed regions. Because of the change in the Moon's axis after about 1 Gyr, it is not clear if water previously accreted onto the Moon could survive but the latter calculation shows that even after 1 Gyr, our mechanism would allow for the accretion of the right amount to explain currently observed water ice amounts on the Moon. One caveat is that it is not certain that the gas released after 1 Gyr evolution would still viscously evolve inwards towards the planets and the Moon because it may well be blown out in a wind-like fashion if the gas density is too low (see Sec.~\ref{evolgas}).

\subsection{Migration of planets}

In our model, the planets end up embedded in a disk with substantial amounts of gas. The resulting planet/gas interactions could therefore lead to the formation of a gap or to the migration of the planet \citep[for a recent review see ][]{2023ASPC..534..685P}. We conducted calculations described in Appendix \ref{appmigr} to compute the drift rate of planets embedded in a gas disk originating from an (exo)asteroid belt. We find that there is probably not enough gas to lead to subsequent drift rates in both type I or II migrations schemes in the Solar System. We find that in some very specific conditions (e.g. massive disk, planets at large distances), the drift rate may become more important and it will be necessary to redo the calculations for each specific extra-solar system.

\subsection{Water delivery in the face of the moon forming impact}

The Moon likely formed as a result of a collision between a young Earth and a Mars-sized object called Theia. The Moon-forming impact happened between $\sim$30 and 200 Myr after the birth of the Solar System, as constrained by the $^{182}$Hf–$^{182}$W chronometer \citep[Hafnium-Tungsten,][]{2009GeCoA..73.5150K} and reinforced by the U/Pb systematics \citep[Uranium-Lead,][]{2010NatGe...3..439R}. Using potentially more precise uranium-lead dating of Apollo 14 zircon fragments, \citet{2017SciA....3E2365B} find that the moon should have formed within 50-60 Myr after the birth of the Solar System, so that most water delivered via our mechanism would have been accreted (at 20-30 Myr) before the Moon-forming impact. 

Interestingly, it is now possible to assess whether accretion of water happened before or after the moon-forming impact by exploring the terrestrial basalt/lunar oxygen isotope difference \citep{2018SciA....4.5928G}. Assuming a nearly complete mixing and isotopic homogenisation in the aftermath of the impact, \citet{2018SciA....4.5928G} find that differences in $\Delta ^{17}$O could possibly indicate a contribution of 5-30\% of material that were accreted after the moon-forming impact. This means that the bulk of Earth's water should have been accreted before the moon-forming impact. This is quite in agreement with our model where most water gets accreted within 30 Myr.

Another hint that water was present early comes from numerical simulations of giant impacts with the Earth. Indeed, simulations show that the presence of an ocean enhances the loss of atmosphere during the giant impact phase in the final steps of planetary formation, which may explain differences in the $^{36}$Ar quantities between the Earth and Venus, being 50 times greater in the latter \citep{2005Natur.433..842G}. More interestingly, simulations show that most water in oceans can be retained despite the violent impacts of this terminal accretion phase or of the moon-forming impact \citep{2003Icar..164..149G,2005Natur.433..842G}.

Contrary to our water delivery mechanism where most water is delivered before the moon-forming impact, most impact-delivery scenarios have more uncertain timings. For instance, for the early instability scenario, because of its unknown timing (between 10-100 Myr), it is not clear whether the moon-forming impact happened before or after it \citep{2024Icar..41416032J}. This may be important to decipher the amount that could have gotten in the mantle as its composition can only be changed before the moon-forming impact. Indeed, after magma ocean solidification there is a decoupling of internal (mantle and crust) and atmospheric reservoirs and water will be mainly retained in the atmosphere. Before the moon-forming impact, the Earth was often in a runaway greenhouse state interacting with a magma ocean surface \citep[e.g.][]{2023Natur.616..306Y} and volatile exchanges between the mantle and the atmosphere were facilitated by liquid–gas interactions at the surface with possible partial volatile retention in the solid or melted portion of the mantle \citep[e.g.][]{2008E&PSL.271..181E}. Shortly after the impact, most volatiles are expected to have partitioned between the atmosphere and the mantle according to their solubility in silicate melt with much of the water going in the mantle, whereas CO, CO$_2$ and most other gases likely go into the atmosphere \citep[e.g.][]{2008E&PSL.271..181E}. Once the mantle has solidified, the mantle no longer convects easily and the heat flow becomes lower thus passing from a runaway greenhouse atmosphere to an atmosphere where water rains down, which facilitates the creation of surface oceans.

\subsection{Current status of water-delivery models and comparison to our new mechanism}

The origin of Earth's water is still to this day an unsolved mystery \citep[][]{2018SSRv..214...47O}. The first issue is whether or not Earth's water has a local primordial origin or was later delivered from other regions of the Solar System, which we explore in the next two sub-sections. Currently, the most popular theory is that Earth formed dry and that water was delivered by a subsequent mechanism \citep[e.g.][]{2009Natur.461.1227A}. There exists a wide variety of later-delivery scenarios, which all imply a delivery by impact with planetesimal (sometimes called left-over planetesimals), asteroidal or cometary material, but can significantly diverge with respect to the exact mechanism causing the delivery as well as its timing. Our aim here is not to present an exhaustive summary of all existing scenarios, but to focus on key features of the most popular ones, which will serve as a reference for comparison with our new disk-delivery mechanism. 

\subsubsection{Local models}
Some studies suggest that water was created locally, or was already there when the local ($\sim 1$ au) primordial bricks that formed the Earth assembled and that it was released later via, e.g. volcanism. For instance, water could have been incorporated into olivine grains at around 1 au through adsorption directly from the gaseous nebula \citep[e.g.][]{2015M&PS...50..578A}. Another study explores the creation of water from the oxidation of an early hydrogen-rich atmosphere by FeO in the magma ocean of the Earth \citep{2008Icar..194...42G}. However, both scenarios would lead to a D/H ratio close to solar, and some additional unlikely processes would be needed to explain current terrestrial D/H (e.g. hydrodynamic escape, fractionation processes). Finally, \citet{2005E&PSL.231....1C} suggest that early in the proto-planetary disk, some water-bearing phyllosilicate dust could drift inwards and be incorporated into planetesimals around 1 au. However, many phyllosilicates found in meteorites appear to have been formed by processes in the parent body after the proto-planetary disk had dissipated. It seems that all of the in situ or very early addition of water scenarios have problems to match data (e.g. D/H ratio). In addition, the first materials that accreted to form the Earth are expected to have been greatly reduced, which would not be the case if they contained a significant amount of water \citep{2008GeCoA..72.1415W}. On the other hand, enstatite chondrite meteorites that have almost identical isotopic compositions to those of the Earth (suggestive of significant contributions to the latter) are drier than the Earth and highly reduced, and it would be difficult to explain if significant water-bearing material was present, unless significant quantities of H atoms could have been released and reacted with oxygen to form water\footnote{It is easy to find O atoms as there is plenty is silicates or minerals in enstatite chondrites but one needs the right conditions of pressure, oxygen fugacity and temperature to get it to stick to H to form water in an efficient way, which is not trivial.} as suggested in \citet{2020Sci...369.1110P}.

\subsubsection{Main impact models}\label{impsec}

Here, we start by listing the main scenarios for water delivery via impacts.

The ``extended feeding zone model'' of water delivery proposes that the giant planets formed near their current locations, and that bodies located at 2.5-4 au, beyond the primordial snowline, and thus icy, can be excited and then accreted onto a forming Earth. Indeed, simulations by \citet[e.g.][]{2006Icar..183..265R} show that after roughly 10 Myr, large bodies with sufficient mass form in the outer regions, which can start scattering smaller bodies by close encounters. It is the end of the oligarchic phase and the outer region is no longer dynamically isolated and can interact with forming planets in the terrestrial region. Bodies in the outer regions can also see their eccentricities pumped up via resonant excitation and secular forcing by Jupiter in addition to mutual close encounters. However, this model is currently disfavored because it fails to create large mass ratios between neighbouring planets like Mars/Earth or Venus/Mercury \citep[e.g.][]{2009Icar..203..644R}. Moreover, it would lead to too much carbonaceous-chondrite-like material accretion in disagreement with data. Indeed, it would lead to volatile element abundances and oxygen isotope ratios of the final planets that are not consistent with Earth's values \citep{2002Natur.416...39D,2012E&PSL.313...56M}. To be consistent, carbonaceous chondrites should have only contributed to about 2\% of the Earth's mass \citep{2012E&PSL.313...56M} but they contribute to 15-20\% in the extended feeding zone model \citep{2006Icar..183..265R}.

The ``Grand Tack model'' proposes that Jupiter migrated inwards in the young proto-planetary disk and then was caught up by Saturn when Jupiter was at 1.5-2 au \citep{2011Natur.475..206W}. This is when the migration of both planets were reversed and they migrated back out until the disk dispersed \citep{2001MNRAS.320L..55M}. In this scenario, the inner part of the asteroid belt is filled with S-type asteroids scattered outwards from the terrestrial planet zone during Jupiter's inward migration, while C-type asteroids have been implanted from outer orbits ($>$ 3 au) during the outward migration. In the process, some scattered asteroids would have impacted the planets and ice-rich asteroids may have been able to deliver water early to those forming planets. In the Grand Tack, water can be delivered early during the inwards and outwards migrations \citep{2023PSJ.....4...32O} or later during the scattering by giant planets of planetesimals located in the outer regions that can last for more than 30 Myr \citep{2014Icar..239...74O}. In this scenario, much less carbonaceous chondritic-like material gets accreted and it remains consistent with carbonaceous chondrites only contributing to about 2\% of the Earth's total mass \citep{2002Natur.416...39D,2012E&PSL.313...56M}. Similar considerations regarding the isotopic ratio of Zn find that only 5\% of C-types are necessary for explaining the volatiles on Earth, while 95\% would come from local enstatite-chondrite like material \citep{2022Icar..38615171S,2022Icar..38615172S}. We also note that recent hydrodynamical simulations at low viscosities\footnote{Indeed, models show that non-ideal MHD effects should lead to non-turbulent proto-planetary disks, even at the surface \citep[e.g.][]{2014prpl.conf..411T}, and observations lead to small $\alpha$ values \citep[e.g. $<10^{-5}$ in][]{2022ApJ...930...11V}.} show that migrating giant planets (like Jupiter and Saturn) go inwards without ever tacking outwards, which may invalidate the Grand Tack scenario \citep{2014ApJ...795L..11P,2023A&A...672A.190G}. One positive aspect of this model is that it explains the small Mars size naturally as its feeding zone gets depleted very early, not allowing Mars to grow significantly. 

{\it The Early Instability model} proposes that a dynamical instability was triggered by the giant planets (after the protoplanetary disk has dissipated, i.e. after a potential Grand Tack for instance) and that it quite likely happened early, between 10 and 100 Myr \citep{2018Icar..311..340C,2019AJ....157...38C}. The instability led to a phase of violent events, which would have strongly depleted the young asteroid belt, leading to impacts on the planets and subsequent water delivery as well as leading to a reduced Mars' feeding zone. It is a very compelling scenario in that it can reproduce features both in the inner and outer regions of the Solar System.

The ``destabilisation by growing planets model'' from \citet{2017Icar..297..134R} proposes that when giant planets are growing, there is a natural mechanism that delivers water to planets. Indeed, they explain that the scattering of planetesimals by destabilizing growing gas giant planets naturally lead to both the implantation of a variety of planetesimals that condensed between 4 and 9 au in the asteroid belt (the precursors of C-types) and water delivery to growing terrestrial planets (because objects are scattered in all directions due to growing planets). 

The ``late veneer'' stage is the addition of mass via impacts to the Earth in the late stage of its formation (i.e. the proto-planetary disk is no longer present), notably after the moon-forming impact. It is expected from measurements of highly siderophile elements in the Earth's mantle that an addition of $5 \times 10^{-3}$ M$_\oplus$ of chondritic material was delivered after the moon formed, i.e. between 30-200 Myr \citep{2009GeCoA..73.5150K} but probably closer to after 50-60 Myr \citep{2017SciA....3E2365B}. It is sometimes argued that the late veneer provided volatile-rich elements but there are many reasons to think that it is actually unlikely that a dominant fraction of Earth's volatiles (including water) could be delivered by the Late Veneer \citep[see more details in][]{2015GMS...212...71M}. Modelling the effect of impacts on the Earth's atmosphere\footnote{using an extended version of the model of \citet{2018MNRAS.479.2649K} used for Trappist-1.}, \citet{2020MNRAS.499.5334S} show that because most planetesimals accreted during the late veneer are left-overs from planet formation, and hence very dry, even assuming that they contain 0.001\% water by mass would only explain 0.25 Earth oceans of water. In addition,  \citet{2020MNRAS.499.5334S} show that the number of impacts needed to explain the Earth's water content would result in too massive an atmosphere compared to current atmospheric mass. They conclude that at most 10\% of water may have been delivered to Earth from impacts (including comets, asteroids and left-over planetesimals) during the late veneer, which is in agreement with geochemical and isotopic arguments also ruling out that a majority of volatiles were delivered during the late veneer \citep{2013GeCoA.105..146H,2015GMS...212...71M}. 

Independently of which impact-scenario is the most efficient, comets have also been suggested as a potential source to deliver water given their obvious ice-rich content \citep[e.g.][]{1995Icar..116..215O}. However, their D/H ratio is diverse and it is often stated to be on average close to twice the terrestrial value \citep{2018SSRv..214...47O}. There has been some hope for Jupiter family comets when 103P/Hartley 2 and 45P/Honda-Mrkos-Pajdušáková were found to have a terrestrial D/H ratio \citep{2011Natur.478..218H,2013ApJ...774L...3L}, but 67P/CG, also a Jupiter-family comet, was found to have a D/H ratio about 3 times terrestrial \citep{2015Sci...347A.387A} and thus more data is needed to look for any potential trend. Another major issue with comets is that the probability of collisions with, e.g. the Earth, is very small and even delivering one Earth ocean of water is thought unlikely \citep{2000M&PS...35.1309M}. Moreover, their $^{15}$N/$^{14}$N ratios are systematically higher than Earth’s \citep{1958GeCoA..14..234J,2016E&PSL.441...91M}, which shows that their contribution should have been less than 10\% overall \citep{2019AJ....157...80C}.

We note that the mechanisms listed above are not mutually exclusive and may happen in parallel (e.g. comet impacts and extended feeding zone model) or one after the other (e.g. destabilisation by growing planets, Grand Tack, and late veneer). There are other models among the impact-like scenarios that are not described here, such as the ``empty primordial asteroid belt model'', ``pebble accretion scenario'', ``convergent migration model'', or ``Jupiter-Saturn resonant chaotic excitation model'' \citep[see the review by][and references therein]{2024arXiv240414982R}. Most models are capable of reproducing the broad strokes of the inner Solar System formation, while also allowing for some water to be delivered.

\subsubsection{Comparison between delivery of water via impacts and accretion from a gas disk}

We note that our water delivery mechanism implies a relatively late arrival of water relative to the dissipation of the proto-planetary disk, with most of it arriving when the Sun is around 25 Myr old, which is consistent with the findings of \citet{2008GeCoA..72.1415W,2015Icar..248...89R} that early material accreting to form Earth must have been highly reduced and the oxidation state increased with time (the most likely oxidant being water). It also means that the Earth can accrete from dry planetesimals (e.g. we do not need the feeding zone to extend beyond 2.5 au) and get its water at a later stage to be consistent with the aforementioned requirement that carbonaceous chondrites only contributed to about 2\% of the Earth's total mass \citep{2002Natur.416...39D,2012E&PSL.313...56M}. Our mechanism would also lead to the correct D/H ratio because water sublimates from C-type asteroids \citep{2012Sci...337..721A} and to the correct $^{15}$N/$^{14}$N ratio because the latter is of chondritic origin and may be already present in the building blocks of Earth \citep{2012E&PSL.313...56M}. Therefore, we do not see any obvious compositional or isotopic reasons that would rule out our delivery mechanism and it even seems that it allows all previous constraints to be matched.
 
We would like to emphasise that we do not rule out the possibility that some of the water could have been transported by impacts before or after the late veneer. However, our scenario comes on top of this, and could be less contingent. Indeed, the impact scenarios described in the previous section (Sect. \ref{impsec}) are mostly based on complex dynamical configurations that may require fine-tuning, such as specific resonances between planets with specific timing to obtain the correct architecture of the solar system and the correct amount of water delivered (e.g. Grand Tack, Early Instability). Our mechanism is essentially agnostic with respect to the complex underlying dynamics that can occur in the system and relies solely on the existence of an early asteroid belt, which does not require fine-tuning.
 
Another advantage of our mechanism is that it seems very efficient because most water ice implanted in the asteroid belt will end up on the planets (half according to our model) with about 1/6$^{\rm th}$ on the Earth (for our fiducial model) whereas one needs many icy objects scattered to be able to have a few collide with the Earth and contribute their water. But we insist that the effectiveness of our mechanism does not rule out impacts. A rigorous comparison between the efficiency of our disk-delivery mechanism and scenarios invoking impacts would be complicated because it needs to be done on a case by case basis, which is beyond the scope of this paper. 

A corollary of our work is that most of the primordial ice in the young asteroid belt will be released before 30 Myr and that, consequently, later impacts will not provide much water to the planets. This potential water depletion of asteroidal projectiles is never taken into account in impact scenarios and could be important in cases where the contributions are late and mostly come from the asteroids (e.g. contribution from asteroids in the late veneer).

Another aspect is that our disk delivery mechanism could relax certain isotopic constraints. For example, if water is mainly supplied by our mechanism, there is a decoupling between the delivery mechanism of volatiles such as N, C or Zn, which may come from chondrites, and the water (some H and O), which could be acquired by means of the disk delivery mechanism. This means that, for example, the D/H ratio constraint does not need to be adjusted within the same scenario as for other volatiles, which may allow for a re-investigation of constraints deduced from isotopic abundances.

\subsection{Testing the abundance of noble gases as a plausible way to discriminate between impact and disk-driven scenarios}\label{sign}

From an observational perspective, it may be hard to distinguish the impact-type scenarios from our mechanism because they both predict that the D/H ratio should be close to that of carbonaceous chondrites, in particular CI and CM types \citep{2012Sci...337..721A}. Could other isotopes, like those of noble gases, be used to discriminate between impacts and our disk-delivery mechanism?

Several sources have been proposed for the Earth's noble gases origin: 1) the implanted solar wind in primordial building blocks of the Earth, 2) accretion of solar nebula gas in the young forming Earth, and 3) a cometary source and/or a chondritic component \citep[for a review see][]{2019AREPS..47..389M}. An interesting point is that some of the noble gases are thought to have been delivered to Earth via impacts of comets or asteroids, which may allow us to make some distinct predictions between the impact scenarios and our mechanism, because the disk-delivery scenario does not lead to any noble gases delivery. Let us explore what we know about each of the noble gases of importance Ne, Ar, Kr, and Xe concerning their origins and whether our mechanism is consistent with noble gases observations on Earth and Mars.

For the light noble gases Ne, and Ar, there are still large uncertainties concerning their origin on the Earth. For instance, \citet{2018GGG....19..979P} note that some atmospheric Neon may have been brought via chondrites and/or comets but it seems more uncertain for Argon because it is hard to decipher whether there is a significant difference between the atmospheric and mantle compositions. Neon isotopes have a solar composition in the Earth's mantle and the origin of this is still debated between either an atmosphere captured from the solar nebula that would get dissolved in a magma ocean, or stellar wind implantation on small grains that later formed the Earth \citep[e.g.][]{2013GChP....2..229M}.

On the other hand, Kr in the Earth's mantle seems to have a chondritic origin mostly originating from early building blocks that assembled into the Earth as measured recently in \citet{2021Natur.600..462}. As for the Krypton isotopic composition of the Earth's atmosphere, it is intermediate between chondritic and solar, which may be explained as a mixture of degassed Kr from the mantle and a late input of cometary material but this is still debated \citep{2021Natur.600..462}. For Xe, it also seems that a cometary contribution is needed to explain the atmospheric isotopic composition on Earth \citep{2017NatCo...815455A} and, potentially, also for the Earth's mantle \citep{2021Natur.600..462}.

Hence, it seems possible to explain the composition of noble gases in the Earth's mantle without late impacts, for example, for Ne (as explained above), and if the building blocks of the Earth have a chondritic composition for Kr or Xe. On the other hand, the Ne and Kr isotopic composition of the Earth's atmosphere cannot be explained by simple outgassing from the mantle, so we probably need a late input (carbonaceous chondrites or comets) after the moon-forming impact\footnote{It is valid to consider that it might be inconsistent to assume different origins for Ne, on the one hand, and Kr and Xe, on the other. However, the solar composition is much richer in Ne than Kr or Xe and it is much less the case for meteorites. Therefore, if it is the dissolution of solar nebula gas in the mantle that explains the solar Ne in the mantle, this process would be less effective for Kr and Xe because they are less soluble than Ne. The same goes for solar wind implantation, as the solar wind is much more enriched in Ne by several orders of magnitude compared with Kr and Xe. So it's possible to have a solar component for Ne and for it to be negligible (and therefore not visible in the current data) for Kr and Xe.} \citep{2019AREPS..47..389M}. 

The mass of impacting comets needed to explain the Ne and Kr atmospheric isotopic ratios is about $3 \times 10^{-6}$ M$_\oplus$ \citep{2013GeCoA.105..146H,2016E&PSL.441...91M,2019AREPS..47..389M}, constituting a minor portion $\sim$ 0.06\% of the probable mass accreted during the late veneer \citep[e.g.][]{2015GMS...212...71M}. It means that the isotopic ratios of atmospheric noble gases can be explained quite easily for the Earth with a few impacting comets during the late veneer. This is because comets are expected to be much richer in noble gases than chondrites and this minor portion of the late veneer does not influence other major volatiles (like the H or N budget). If instead there are no cometary impacts and one assumes that only carbonaceous chondrites get accreted, the noble gases present in the atmosphere are not well fitted and it would lead to significant overabundances in the terrestrial budget of platinum group elements \citep{2019AREPS..47..389M}.

Some recent results on noble gases on Mars have also been obtained by measuring the isotopic composition of the martian meteorite Chassigny, representative of the planet’s interior \citep{2022Sci...377..320P}. They find chondritic krypton isotope ratios, which implies early incorporation of chondritic volatiles. But the atmosphere of Mars has different (solar-like) krypton composition, indicating that it is not a by-product of magma ocean outgassing and may come from early accretion of solar nebula gas after formation of the mantle. Hence, no late veneer impacts are needed to explain Mars's Kr data. 


As a consequence, noble gases cannot be used to discriminate between our disk-driven mechanism and impact-driven scenarios, because only a few late impacts are needed to explain noble gases atmospheric data, which should happen whether or not there is a disk-delivery of water in addition to it. It is thus required to look for more tracers that would be able to differentiate the impact-delivery from our mechanism. We suggest that the best test to know whether our scenario is valid and quantify it better may be to look for water gas disks in extrasolar systems as we explain in further detail in Sect. \ref{obswat}. It is well possible that after discovering the ubiquity of CO gas disks at the debris disk stage in exo-Kuiper belts \citep[e.g.][]{2017ApJ...849..123M}, we discover the ubiquity of water gas disks at the same stage in exo-asteroid belts.

\subsection{Collisional evolution of solid bodies in the solar system}

If the asteroid belt starts with a low mass of $4 \times 10^{-4}$ M$_\oplus$, then the largest bodies do not have time to collisionally evolve significantly; namely, the change in mass for the most massive asteroids is negligible. Indeed, using eq.31 in \citet{2008ApJ...673.1123L}, we can compute that for solid bodies between 100 km and 1000 km with a steep size distribution in -4.5, the collisional lifetime of a 200 km asteroid in the middle of the main belt ($\sim 2.65$ au) is $\sim 50$ Gyr, i.e. much larger than the age of the Solar System. For a belt of 0.1 M$_\oplus$ similar to that of the second scenario, the collisional lifetime gets down to 0.2 Gyr but it is still much longer than the 50 Myr when we deplete most of the mass in scenario 2 to mimic an early instability. We conclude that the collisional evolution in the main belt should not affect our results significantly because most of the ice mass is in large bodies $> 100$ km.

\subsection{Ejection of dust caused by water ice sublimation}

Based on 67P/CG in situ analysis, it is clear that, for comets, the sublimation of gas releases dust along with it. The dust-to-water mass ratio for escaping material is around 1.5 \citep[likely between 0.86 and 2.6,][]{2020SSRv..216...44C}. The size distribution index of cometary dust that comes off ranges from -3.5 to -4 \citep{2023arXiv230503417E} with a minimum size around 10 microns \citep[with smaller sizes existing but not dominating the cross-section,][]{2018AJ....156..237M,2019A&A...630A..24G} and a maximum size between 1 mm and 1 m that depends on the gas sublimation rate \citep{2020FrP.....8..227M} because more activity can allow to lift larger grains. 

However, the physics of ejection is expected to be much different for the large ($>$100 km) asteroids that are probably the main ice reservoir for our disk-delivery mechanism, as compared to comets that are typically 1 to a few tens of km (e.g. $\sim 3$ km for 67P/CG) from which material escape more easily. For these large 100-1000 km asteroids, the escape velocity is 30-300 times greater than for 67P/CG, and the dust mass that can be dragged away by sublimating gas is thus necessarily greatly reduced. Indeed, using the cometary model of \citet{2018Icar..312..121Z} with standard values for large asteroids, we find that the largest grain size that can escape, dragged along with gas, is around 1 micron, while all larger dust falls back onto the asteroid after ejection (hence paradoxically no bright dusty tails are expected for those large bodies). These small micron-sized grains are not expected to be very abundant on comets \citep{1997Icar..127..319C} and may not be abundant on large asteroids \citep{2013Icar..223..479G}. Overall, we expect that it will not be a large mass of dust that is released from large asteroids because most mass is contained in larger fragments that do not escape. 

If the dust ejection level is reduced for large asteroids, there may still be activity releasing dust for bodies smaller than 100 km. Due to the flatter size distribution of asteroids between 20-100 km ($q_{\rm med}=1.2$), the mass as a function of body size varies greatly in this range. The initial ice mass in bodies smaller than 30 km is $\sim 10^{-6}$ M$_\oplus$ for scenario 1 and $\sim 2 \times 10^{-4}$ M$_\oplus$ for scenario 2. To gauge how much dust mass could be removed from asteroids during water sublimation from bodies smaller than 30 km, we  assume a dust-to-water mass ratio of 1.5 so that we find that the activity could release a dust mass of $\sim 1.5 \times 10^{-6}$ M$_\oplus$ for scenario 1 and $\sim 3 \times 10^{-4}$ M$_\oplus$ for scenario 2. Most of it will be released between 20-30 Myr during the surge in luminosity and its subsequent evolution will depend on the grain sizes. Because of the steep size distribution of dust released during activity (based on cometary results, see above), most grains will have a size close to the minimum size around 10 microns for which the ratio of radiation pressure to gravitational forces $\beta_{\rm rp} \sim 10^{-2}$ \citep{1979Icar...40....1B}. This dust will thus spiral inwards from the main belt to the Sun due to Poynting Robertson drag over timescales of $2.5 \times 10^5$ yr  \citep{2005A&A...433.1007W}. Assuming that the dust is released over 10 Myr (between 20 and 30 Myr after the birth of the Sun) the dust mass that will be present in the solar system at any moment during the surge in luminosity is thus a factor 0.025 smaller and closer to $4 \times 10^{-8}$ M$_\oplus$ for scenario 1 and $\sim 8 \times 10^{-6}$ M$_\oplus$ for scenario 2. We note that this component is far from negligible as it is close to typical dust masses measured in exo-asteroid belts \citep{2014ApJS..211...25C} and it should be accounted for in models trying to predict the dust quantity in warm belts but also when trying to compute the amount of warm and hot dust, called exozodis, in a specific system \citep{2017AstRv..13...69K}.
 



\subsection{Consequences for extrasolar systems: A universal mechanism}\label{universal}

Our disk-based water delivery mechanism seems more universal than impact-type scenarios, because it is less contingent and can occur regardless of the complex dynamical history of the system. It seems universal across a large range of planetary systems (including a large variety of stellar types) because asteroid belts in younger proto-planetary disks are expected to be colder than in debris disks (see the model below), leading to a snowline that always moves outwards when the primordial gas dissipates and thus allowing degassing of water ice that can now sublimate. To better quantify this displacement of the snowline, we compare a simple model of temperature evolution of a young proto-planetary disk to a black body temperature (i.e. the temperature of solids after dissipation of the young disk). 

\begin{figure}
   \centering
   \includegraphics[width=9.cm]{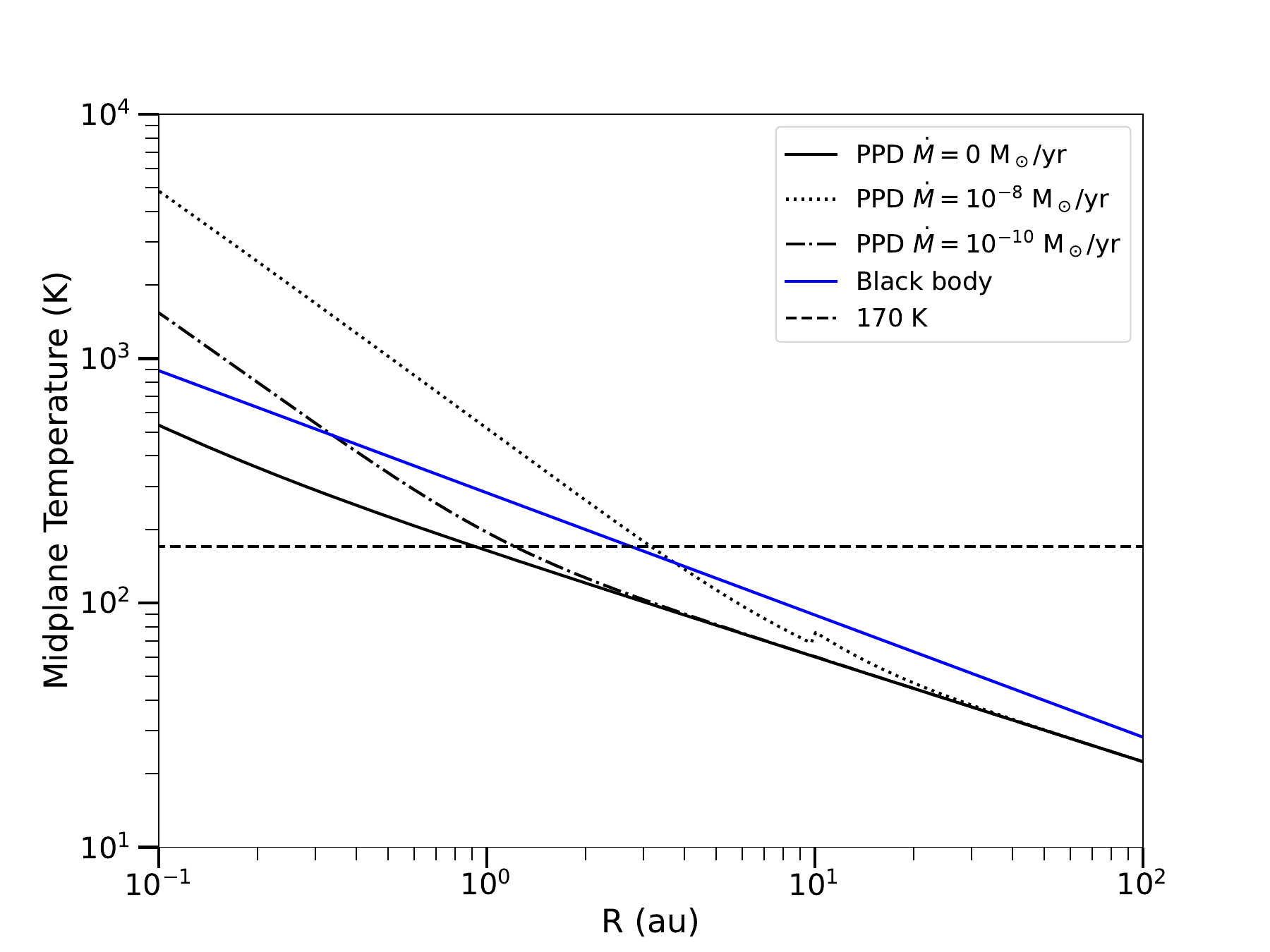}
   \caption{\label{figtempppd} Temperature as a function of radial distance to the star in a proto-planetary disk (black) Vs black body temperature (blue). For the proto-planetary disk case, the different linestyles represent different accretion rates onto the star (0 for the solid line, $10^{-8}$ M$_\odot$/yr for the dotted line, and $10^{-10}$ M$_\odot$/yr for the dash-dotted line), which corresponds to different timescales (e.g. roughly 1-2 Myr for $10^{-8}$ M$_\odot$/yr and later for lower accretion rates). The dashed horizontal line shows a temperature of 170 K, close to the sublimation temperature of water ice. We see that the snowline moves initially inwards when the disk accretion rate gets lower and then outwards when the disk fully dissipates and the temperature of asteroids gets close to a black body temperature.}
\end{figure}

For an optically thick proto-planetary disk, where the radiation from the star impinges at an angle $\phi(r)$, \citet{2001ApJ...560..957D} show that the midplane temperature profile due to stellar irradition can be simplified to

\begin{equation}\label{tempirr}
T_{\rm mid,irr}=\left(\frac{\phi}{2} \right)^{1/4} \left(\frac{R_\star}{r} \right)^{1/2} T_\star,
\end{equation}

\noindent where $R_\star$ is the radius of the Sun, $T_\star$ the effective temperature of the Sun and $\phi(r)=0.4 R_\star/r+0.05r^{2/7}$ \citep[e.g.][]{1997ApJ...490..368C,2000ApJ...528..995S}. The midplane temperature due to accretional heating can be derived as \citep[e.g.][]{2006ApJ...640.1115L}:

\begin{equation}\label{tempacc}
T_{\rm mid,acc}=\left[\frac{3}{4} \left(\tau_R+\frac{2}{3} \right) \right]^{1/4} T_{\rm eff,acc},
\end{equation}

\noindent where $\tau_R$ is the Rosseland optical depth and $T_{\rm eff,acc}$ is the effective temperature corresponding to the total flux released by accretional heating with \citep[e.g.][]{2006ApJ...640.1115L}:

 \begin{equation}
 \sigma T_{\rm eff,acc}^4=\frac{3}{8 \pi} \frac{GM_\star \dot{M}_{\rm acc}}{r^3} \left(1-\sqrt{\frac{R_\star}{r}} \right),
\end{equation}

\noindent where $\dot{M}_{\rm acc}$ is the accretion rate onto the star of mass $M_\star$. Accounting for both the irradiation from the central star and accretional heating, the final midplane temperature can be calculated as

 \begin{equation}
 T_{\rm mid}^4=T_{\rm mid,irr}^4+T_{\rm mid,acc}^4,
\end{equation}

\noindent which we plot in Fig.~\ref{figtempppd} (black). We note that the Rosseland opacity $\kappa_R$ near the sublimation temperature is around 1 m$^2$/kg \citep{2008ApJ...679..797J,2016A&A...590A..60B}, which we use as a simplification in our model (because the dependence of the temperature scales as $\kappa^{1/4}$) using: $\tau_R= \kappa_R \Sigma_{\rm MMSN}$ with the minimum mass solar nebula density equal to $\Sigma_{\rm MMSN}=1.7 \times 10^4 (r/1{\rm au})^{-3/2}$ kg/m$^2$. However, viscous evolution quickly reduces the initial MMSN surface density in the inner 10 au and the radial slope of the surface density evolves to -1 \citep[e.g.][]{2019A&A...624A..93B}. To be more accurate, we use the MMSN beyond 10 au and reduce it to $2000 (r/1{\rm au})^{-1}$ kg/m$^2$ within 10 au, which is the MMSN value after 1-2 Myr. This is in line with an $\dot{M}_{\rm acc}$ value of $10^{-8}$ M$_\odot$/yr \citep[e.g.][]{2019A&A...624A..93B}.

Figure~\ref{figtempppd} shows the results of our temperature evolution model. We can see that accretional heating dominates when the disk is still young with a strong accretion rate onto the star (dotted line where $\dot{M}_{\rm acc}=10^{-8}$ M$_\odot$/yr), and pushes the snowline further out at around 3 au (similar to the black body temperature in blue). But when viscous heating diminishes and that the gas disk gets to lower masses, the midplane temperature decreases and the snowline (the thin horizontal solid line shows a temperature of 170 K close to the sublimation temperature) moves inwards to $\sim 1$ au. When the disk dissipates the temperature increases to a value close to the black body temperature (in blue, Eq~\ref{Tbb}) which is higher and the frozen ice can now sublimate (the snowline moves back out). We used realistic values of $R_\star$, $T_\star$, and $L_\star$ from our solar evolution model (see Sect. \ref{sun}) and tested for two different ages of the Sun, 2 and 10 Myr, which led to very similar results and we only show the former in Fig.~\ref{figtempppd}. Note that this is an approximate model but it shows results that are not at odds with more complex self-consistent models \citep[e.g.][]{2016A&A...590A..60B}. 

Comparing $T_{\rm mid,irr}$ with $T_{\rm bb}$ (Eq~\ref{Tbb}) is instructive, as it can be used to explore this behaviour for different stellar types. Indeed, after using that $4\pi R_\star^2 \sigma T_\star^4=L_\star$ and rearranging, we obtain $T_{\rm mid,irr}/T_{\rm bb}=(2 \phi(r)/(1-A))^{1/4}$, with $\phi$, defined above, being dominated by the term $0.05 r^{2/7}$ at exo-asteroid belt locations; hence, it has values close to 0.05-0.1, making the midplane temperature due to irradiation around twice lower than the black body temperature for all stellar types. We note that this explains why all asteroid belts located at the right position (i.e. with temperatures that can reach $>$150 K once the gas has dissipated to allow for sublimation to start) should initially sublimate but the sublimation rate will evolve over long timescales in different ways, depending on the temporal evolution of the stellar luminosity. For instance, in the case of the Sun, most of the gas is released when its luminosity increases around 20 Myr and this behaviour may become even more complex for more massive stars having more phases of luminosity surge \citep[e.g.][]{1967ApJ...147..624I} that may lead to several phases of water delivery.

The conclusion is that because of the opacity of young massive proto-planetary disks, asteroid belt locations are colder before the disk dissipates and water ice can accumulate. But when the young disk dissipates, icy asteroids can suddenly find themselves interior to the snowline and start sublimating, which creates a gas disk that can viscously expand to the planets and potentially deliver large quantities of water onto them. This means that the prerequisites for our scenario to unfold should be generically met (at least qualitatively) in young asteroid belts. It seems therefore essential to start looking for those water disks in extrasolar systems, especially because warm belts seem to be located close to the primordial snowlines and the same mechanism as explained in this section should be at work in extrasolar systems \citep{2011ApJ...730L..29M,2017ApJ...845..120B}. In Sect. \ref{obswat}, we investigate whether we could detect these water gas disks with current technology.

\subsection{Observations of water in extrasolar systems}\label{obswat}
\subsubsection{ALMA}
Water gas can be targeted with ALMA and explored at high resolution to look for the snowline position in proto-planetary disks as shown recently in \citet{2024NatAs.tmp...49F}. There are at least three transitions of water of interest in bands 5 and 7 that are p-H$_2$O $3_{13}-2_{20}$ and $5_{15}-4_{22}$, at 183.31 and 325.15 GHz, respectively, and o-H$_2$O $10_{29}-9_{36}$ at 321.22 GHz, where two lines are of para-water and one line of ortho-water.

It is beyond the scope of this paper to fully explore the parameter space of plausible detections with ALMA. Here, we focus here on an illustrative example which is the promising system around the star HD 69830 that hosts a warm belt at $\sim 1$ au \citep{2009A&A...503..265S} in which water ice has been detected \citep{2007ApJ...658..584L}. 

\begin{figure}
   \centering
   \includegraphics[width=9.cm]{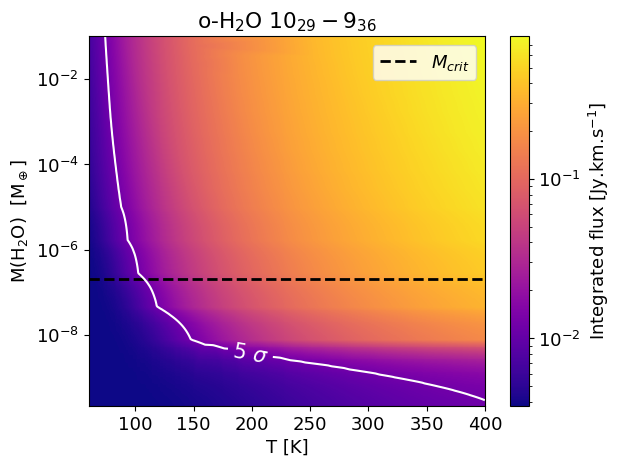}
   \caption{\label{figalma} Alma observation predictions to target water in HD 69830 showing the integrated flux in Jy.km.s$^{-1}$ for the o-H2O $10_{29}-9_{36}$ transition at 321.22 GHz as a function of the temperature and the quantity of water. The black dashed line indicates the critical mass level when water starts self-shielding. The level at 0.012 Jy.km.s$^{-1}$ is for a signal-to-noise ratio of 5 represented by the solid white line labelled 5$\sigma$. This line is rather sensitive to low amounts of water and may allow for the detection of the first water gas disk in the near future.}
\end{figure}

We ran several simulations with the radiative transfer code RADMC3D \citep{2012ascl.soft02015D} to estimate the integrated flux of water lines in bands 5 and 7 as a function of the total water content and gas temperature. Our simulations show that the water transition line o-H$_2$O $10_{29}-9_{36}$ at 321.22 GHz (band 7) is the most sensitive to detect gaseous water released from the icy belt in HD 69830. We assume LTE in our simulations. Fig.~\ref{figalma} shows the integrated line flux for this transition in the water mass Vs gas temperature parameter space. This shows that if the gas temperature is greater than 100 K (which is likely given the dust temperature of $\sim$ 275 K in the belt), we could readily detect a total mass of water gas above that needed for water shielding (i.e. above the black dashed line). This is promising because the dust mass in the warm belt of HD 69830 is about 10 times that of the current asteroid belt, and using Fig.~\ref{figgasmass}, it indicates that the gas disk around HD 69830 may indeed be more massive than $10^{-6}$ M$_\oplus$. The comparison is appropriate given that HD 69830 is a K0V star \citep{2015ApJ...800..115T} with a belt close to 1 au but dedicated simulations would be needed to make accurate predictions. The black dashed line shows that a gas mass above $2 \times 10^{-7}$ M$_\oplus$ ensures that water will be shielded rather than photo-dissociated (see Sect. \ref{secshiel}). We can only use water lines for masses above this line because we need water molecules and not their photo-dissociation products (H and O) to probe the water lines. The estimated gas mass of $10^{-6}$ M$_\oplus$ thus shows that the gas will be composed of water and the total water mass predicted in HD 69830 is high enough for detections with ALMA even in the case of higher $\alpha$ values.


We note that it is intriguing that this system still shows traces of water ice given its age (3-10 Gyr) but it may be explained by several factors: 1) the star is less massive than our Sun \citep[$\sim$ 0.86 M$_\odot$,][]{2006Natur.441..305L} and its luminosity is about 0.6 L$_\odot$, meaning that sublimation will happen on longer timescales for a given radial position to the star, or 2) the stellar evolution luminosity profile may differ and a surge in luminosity may happen at different times, or 3) the initial water ice-to-rock fraction or albedo may have been higher.

This system is all the more interesting because it hosts 3 known planets, the most distant of which, HD 69830 d (at $\sim$0.68 au with a mass greater than 18 M$_\oplus$), may be in the habitable zone with an orbit of $\sim$200 days \citep{2006Natur.441..305L}. The possibility that there is enough water left to be efficiently accreted by these planets is a fascinating one. We therefore suggest that ALMA could open a new avenue as it can be used to target water gas down to low levels. ALMA would even have the capability to resolve the gas disk at $\sim$1 au by going to high resolution (e.g. C-7 configuration or more extended) because the system is located at 12.6 pc, which could be an opportunity to refine the gas models and look for gaps due to planets in the profile.

\subsubsection{JWST (exploratory work)}
Observations with the JWST, though at lower spectral and spatial resolutions, would also be insightful. Indeed, water lines are present everywhere in the mid-infrared. \citet{2023ApJ...959L..25X} have for example shown that, for the young (1-3 Myr) class II disk Sz 114 the JWST can detect myriads of water lines  from the hot bending rovibrational band at $\sim$6.6 $\mu$m to the cooler pure rotational lines at $\sim 25$ $\mu$m with, e.g. MIRI. {\it Spitzer} has targeted warm belts in the mid-IR and though it may have detected water ice, it did not find any water gas lines \citep[e.g.][]{2007ApJ...658..584L}. However, the JWST has a collecting surface area 58 times larger than that of {\it Spitzer}, which may allow for  those faint gas lines to be detected.


There are water lines that could be targeted at small wavelength with NIRSpec in addition to the longer wavelengths with MIRI (e.g. MRS) but the full study of the best observing strategy and comparison to ALMA are out of the scope of this paper. We leave the JWST predictions for future work but note that the HD 69830 system may be the most obvious target because of the potential presence of water ice in an asteroid belt. However, it may be interesting to make predictions for a wider variety of systems and target them with the JWST in the near future as well as to look into archival data.

\subsection{Consequences on the habitability of exo-Earths}\label{habwat}

Liquid water is thought to be a necessary ingredient for the development of life on a planet \citep[e.g.][]{2014PNAS..11112628M}. Exo-Earths, which are planets with a size and temperature similar to Earth's, have a similar problem to our blue planet in that they probably form dry and need exogenous sources to provide water. Impact-type scenarios often need a complex dynamical history requiring the right resonances or planetary migration rates to scatter a certain amount of icy bodies from outer planetary regions onto the forming exo-Earths. This could mean that it is difficult to deliver water to exo-Earths and far from generic. However, our disk-delivery mechanism is less contingent and can occur in all systems with exo-asteroid belts, meaning that it may not be so difficult to deliver water to young exo-Earths in formation.

If our disk-delivery mechanism is dominant over the other scenarios, this means that future missions targeting exo-Earths may need to search for systems with exo-asteroid belts that may contain exozodiacal dust (because of dust produced by collisions in the exo-asteroid belt and of activity due to sublimation) that could add noise to the signal and prevent the detection of exo-Earths using high contrast direct imaging \citep[e.g.][]{2012PASP..124..799R}. It is therefore important to swiftly understand how water is transported and whether we should target exo-Earths close to dusty exo-asteroid belts to optimise our chances of having water on the surfaces of the exo-Earths in the quest for detecting life. This point has profond implications as it could change the way we want to approach exo-Earth detections from an observational standpoint in the coming decades.

\section{Conclusions}

There are many reasons to think that young asteroids were made up of a fraction of water ice (e.g. most C-type asteroids are hydrated, which could be the signature of an earlier icy component) that sublimated over time to lead to the current situation where most asteroids appear to be ice-free. In this article, we explore the effect of these young icy asteroids sublimating over time and show that this offers a new mechanism capable of transporting water to the inner planets in the Solar System. It may even be universal across all exo-planetary systems with respect to habitable-zone planets.

We have developed a new model that allows us to follow the sublimation of these young icy asteroids over time, taking into account the evolution of the luminosity of the young Sun. Here, we show that a disk of gas filled with water can be rapidly created in the inner region of the Solar System. We follow the viscous evolution of this disk and find that it propagates inwards and outwards towards the surrounding planets. The gas disk can last for hundreds of Myr because it is predominantly sustained by the sublimation of the largest objects in the belt, which can retain ices for considerable amounts of time. We find that the mass of the gas disk reaches a maximum when the Sun's luminosity reaches $\sim 1$ L$_\odot$, around 20-30 Myr after its birth. This is when most of the gas is released from the icy asteroids, which could be an ideal spot to try and detect these new types of aqueous gas disks in extrasolar systems.

In our model, we quantify the amount of water that can be accreted onto the terrestrial planets (Mercury, Venus, Earth, and Mars) and the Moon with our viscous spreading transport mechanism for two distinct scenarios: 1) the asteroid belt was initially massive, as indicated by the MMSN model, and depleted early at $\sim 50$ Myr and 2) the asteroid belt started with its current mass. We find that in the case of an initially massive belt, our water transport mechanism can explain the water content data for the terrestrial planets and the Moon in our solar system without the need of any additional source. For example, for the Earth, we find that we can easily transport ten terrestrial oceans of water, which corresponds to the total amount of water in the hydrosphere and mantle measured in the Earth. We note that our model gives an upper limit on the initial asteroid belt mass of $\sim 0.1 \times \frac{f_{\rm ice}}{0.2}$ M$_\oplus$, where $f_{\rm ice}$ is the fraction of ice-to-solid on initial asteroids. This is because a more massive belt would lead to too much water accreted onto the Earth.

Our mechanism would also be in agreement with the Earth's D/H ratio measurements, because most of the water would initially have come from the sublimation of C-type asteroids, which have a terrestrial D/H ratio. Other important constraints such as the $^{15}$N/$^{14}$N ratio, the late veneer and noble gases would not be affected by our mechanism. Our new mechanism could have an influence on the results of works attempting to fit the isotopic ratios of the various volatiles together. In fact, we propose that there is a possible decorrelation of the D/H ratio (which would be set by disk delivery) in relation to the other volatiles that could be contributed by the building blocks or subsequent impacts, which could change the conclusions of many studies.



We find that this new mechanism of disk-driven water distribution is inevitable and may in fact be universal in systems with exo-asteroid belts around a wide range of stellar types. The main requirements to trigger it are indeed that (exo)asteroids form in an initially colder environment allowing them to retain water ice and that later on the snowline moves outwards, leading to the sublimation of these ices and the creation of a viscous gas disk. These requirements should be met in most young planetary systems, where the opacity of the proto-planetary disk should maintain low-enough temperatures for the first $\sim$ 5 Myr, followed by a temperature increase after the dissipation of this primordial disk, allowing ices to sublimate. This sublimation process then naturally creates a water-vapour secondary disk that inevitably spreads towards the inner region and deliver water to exoplanets.


The best observational constraint for better quantifying the importance of our mechanism for the Solar System and exoplanetary systems would be able to detect disks of aqueous gas in extrasolar systems. We estimate that ALMA (and potentially JWST) could detect these disks, provided that they are targeted at the right moment when the water gas disk is not overly diluted. More precise predictions for different stellar types would be useful to increase the probability of detecting these aqueous gas disks in the coming years.

\begin{acknowledgements}
This paper is dedicated to Grégory. We thank the referee for their helpful report. We wish to thank Louis Manchon for providing a state-of-the-art model of the solar evolution and some description about it. We also thank Kévin Baillié, Bruno Bézard, Benjamin Charnay, Sandrine Péron, Laurette Piani, and Vladimir Zakharov for interesting discussions.
\end{acknowledgements}

\appendix

\section{Migration rates of planets in gas disks originating in asteroid belts}\label{appmigr}

The speed at which a planet migrates depends on its ability to form a substantial gap: a planet producing a small gap (i.e. with a density within the gap which is less than half of the density outside) can rapidly migrate in a Type I migration scheme, whereas a gap-opening planet migrates more slowly in a Type II migration scheme. \cite{2015MNRAS.448..994K} show that the gap depth - the ratio of the minimum density in the gap $\Sigma_{\rm min}$ to the unperturbed density out of the gap $\Sigma_{\rm un,p}$ - can be written as

\begin{equation}
    \frac{\Sigma_{\rm min}}{\Sigma_{\rm un,p}} = \frac{1}{1 + 0.04 K}
    \label{eq_gapdeth}
,\end{equation}

\noindent where,
\begin{equation}
    K = \left( \frac{M_p}{M_\star} \right)^2 \left( \frac{H_p}{r_p} \right)^{-5} \alpha^{-1} = \left( \frac{M_p}{M_\star} \right)^2 \left( \frac{r_p k_B T_p}{GM_\star\mu m_{\rm p}} \right)^{-5/2} \alpha^{-1}
    \label{eq_K}
,\end{equation}
\noindent with $M_p$ and $M_\star$ being the mass of the planet and the star respectively, $H_p = c_s(r_p)/\Omega_p$ (with $\Omega_p=\Omega(r_p)$) is the disk scale height at the planet position $r_p$ and $\alpha$ is the turbulent viscosity parameter.

It is possible to estimate the total torque and the resulting migration speed from this gap depth. In the small gap cases, the total torque that the planet feels can be approximated by (for $K < 5$) $\Gamma = -c\Gamma_0(r_p)/(1 +0.04K) \approx - c\Gamma_0 (r_p) $ \citep{2018ApJ...861..140K} where $\Gamma_0(r) = (M_p/M_\star)^2 (H_p/r)^{-2} \Sigma_{\rm un} r^4 \Omega_p^2 $ is the normalizing torque at distance $r$ of an unperturbed disk of surface density $\Sigma_{\rm un}(r)$ and $c$ is a numerical constant varying between 1 and 3, depending on $\alpha$ \citep{2018ApJ...861..140K}. This torque leads to the migration of the planet and the new planet position can be estimated by \citep{1996ApJ...460..832T}

\begin{equation}
\frac{d r_p}{d t}=\frac{2 \Gamma}{M_p \Omega_p r_p},
\end{equation}

\noindent which leads to the following drift rate in the case of small gaps:

\begin{multline}
     \frac{dr_p}{dt} \approx - 3\times10^{-6} \; \left(\frac{M_\star}{1 \rm M_\odot}\right)^{-1/2} \left(\frac{M_p}{1 \rm M_\oplus}\right) \left(\frac{\mu}{18}\right) \left(\frac{\Sigma_{un}}{10^{-3} \rm \, kg/m^2}\right) \\ \left( \frac{r_p}{\rm 1 au} \right)^{1/2} \left( \frac{T_p}{278.3 \rm K} \right)^{-1} [\rm au/Myr].
     \label{eq_smallK}
\end{multline} 

On the other hand, for gap-opening planets, the total torque applied on the planet can be written as $\Gamma \approx -c\Gamma_0(r_p)/(0.04K)$ \citep{2018ApJ...861..140K} leading to:
\begin{multline}
      \frac{dr_p}{dt} \approx - 2.6 \times 10^{-7} \left( \frac{M_\star}{1 \rm M_\odot} \right)^{-1} \left( \frac{M_p}{1 \rm M_{\rm Jup}} \right)^{-1} \left( \frac{\mu}{18}\right)^{-3/2} \\ \left( \frac{\Sigma_{un}}{10^{-4} \rm \, kg/m^2} \right) \left( \frac{r_p}{5.2 \rm au}\right)^3  \left( \frac{T_p}{122 \rm K}\right)^{3/2} \left(\frac{\alpha}{10^{-2}}\right) [\rm au/Myr].
     \label{eq_largeK}
\end{multline}

In our disk with $\mu = 18$, $\alpha = 10^{-2}$ in the most massive disk (i.e. with $\Sigma_{\rm un,p} \sim 10^{-3}$ kg/m$^2$, see Fig.~\ref{figsigma} bottom), the terrestrial planets hardly open gaps (see Table \ref{tab:gaps}). However, for smaller values of $\alpha$, for instance, $10^{-3}$, the gaps for the Earth and Venus would become more important with $\Sigma_{\rm min}/\Sigma_{\rm un,p} \sim 0.4$. For the largest planets, such as Jupiter, deeper gaps are opened (Table \ref{tab:gaps}). For these cases, the resulting type II migration is highly inefficient, barely pushing the giant. If Jupiter indeed creates such a deep gap, we can wonder whether the asteroid-belt produced gas would manage to cross its orbit outward towards Saturn, though it does not affect our conclusions for the accretion of water on terrestrial planets. Indeed, it is possible that due to the presence of Jupiter, the gas piles up at the inner edge of its gap, preventing this water-rich gas from reaching the outer planets. We note that the presence of a gap or a sharp density drop could then be used as a signature for the presence of a planet \citep{bc}.

As for the drift rates of planets, Eqs.~\ref{eq_smallK} and \ref{eq_largeK} show that there is not enough gas in the disk interacting with planets for them to be affected and significantly migrate in this secondary water gas disk in the case of both small and large gaps. We note that for extrasolar systems having even more massive disks for extended periods of time, or less massive gap-opening planets (e.g. sub-Neptunes) or for planets at larger distances from their stars (the migration rate scales as $r_p^3$ for gap-opening planets), the drift rate will be more important and these calculations would need to be adjusted to consider the potential migration of planets in subsequent studies.

\begin{table}[t]            
\centering                          
\caption{K values, gap depths and migrating distances in 10 Myrs for different Solar System planets assuming $\Sigma=10^{-3}$ kg/m$^2$ for terrestrial planets and $\Sigma=10^{-4}$ kg/m$^2$ for giants and the other parameter values are the same as fiducial values used in Eqs.~\ref{eq_smallK} and \ref{eq_largeK} (e.g. $\alpha=0.01$).} 
\begin{tabular}{c c c c}        
\hline             
        & $K$ & $\Sigma_{\rm min}/\Sigma_{\rm un,p}$ & $\Delta r_p$ [au]  \\
\hline
Mercury  & 0.035 & 0.999 & -$6.5  \times 10^{-7}$ \\
Venus    & 3.779 & 0.869 & -$1.7 \times 10^{-5}$   \\
Earth    & 3.643 & 0.873 & -$3.0 \times 10^{-5}$  \\
Mars     & 0.025 & 0.999 & -$4.7 \times 10^{-6}$  \\

 \hline                  
Jupiter  & 46885 & $5.3\times 10^{-4}$ & -$2.6 \times 10^{-6}$  \\
Saturn   & 1952  & 0.0126              & -$3.5 \times 10^{-5}$  \\
  
\hline
\end{tabular}
\label{tab:gaps}
\end{table}

\section{Details on the Sun's model used in this paper}\label{appsun}

The Solar evolution model used for this work was computed with the \cesamxx\footnote{\url{https://www.ias.u-psud.fr/cesam2k20/}} stellar evolution code \citep{1997A&AS..124..597M,2008Ap&SS.316...61M,2013A&A...549A..74M,2024A&A...687A.146M}. Stellar structure equations (for the pressure, radius, temperature and luminosity as functions of the mass coordinate) are solved using a collocation method where solutions are represented as a linear combination of piecewise polynomials, projected over a B-Spline basis. The present model uses \citet{2009ARA&A..47..481A}'s determination of the solar chemical composition, with opacity tables from the OPAL team \citep{1992ApJS...79..507R,1996ApJ...464..943I}, adapted to these abundances. The equation of state (\eos) uses tables from the OPAL5 \eos \citep{2002ApJ...576.1064R}. The nuclear reaction rates follow the compilations from NACRE \citep{2006AIPC..831...26A} except for the ${}^{14}{\rm N(p},\gamma){}^{15}{\rm O}$ reaction where the LUNA facility has been used \citep{2018PrPNP..98...55B}. The convection is modelled with the mixing-length theory formalism under the formulation of \citet{1965ApJ...142..841H}, taking into account the optical thickness of the convective bubble. The atmosphere is retrieved using the Hopf function $q(\tau)$ given in \citet{2014tsa..book.....H}. Effects of rotation and diffusion are not considered.

 The solar model was calibrated using the  program Optimal Stellar Model\footnote{\url{https://pypi.org/project/osm/}} (OSM), interfaced with \cesamxx, that implements a Levenberg-Marquardt algorithm. Initial helium content $Y_0$ and the mixing-length parameter $\alpha_{\rm MLT}$ were tuned in order to retrieve, at solar age, the luminosity $1\,L_\odot$ \citep[$L_\odot = 3.828\times10^{26} W$;][]{2015arXiv151007674M} and the effective temperature $5772\,\K$ of the Sun, with a maximum error of respectively $10^{-5}$ and $1\,\K$. For this model, optimal values are found to be $Y_0 = 0.2553$ and $\alpha_{\rm MLT} = 1.631$.

\label{lastpage}

\end{document}